\documentclass[english,prd,twocolumn,superscriptaddress,showpacs,preprintnumbers,nofootinbib]{revtex4-1}

\usepackage{graphicx}
\usepackage{dcolumn}
\usepackage{bm}

\usepackage{amsmath}
\usepackage{amssymb}
\usepackage{pifont}
\usepackage{tabularx} 
\usepackage{etoolbox}
\usepackage{booktabs} 
\usepackage{float}
\usepackage[dvipsnames, svgnames]{xcolor}

\usepackage{hyperref}

\hypersetup{
    colorlinks=true,
    linkcolor=blue,
    citecolor=blue,
    urlcolor=blue
}

\begin{document}

\title{Low-Energy Calibration of SuperCDMS HVeV Cryogenic Silicon Calorimeters Using Compton Steps}

\author{M.F.~Albakry} \affiliation{Department of Physics \& Astronomy, University of British Columbia, Vancouver, BC V6T 1Z1, Canada}\affiliation{TRIUMF, Vancouver, BC V6T 2A3, Canada}
\author{I.~Alkhatib} \affiliation{Department of Physics, University of Toronto, Toronto, ON M5S 1A7, Canada}
\author{D.~Alonso-González} 
    \affiliation{Departamento de F\'{\i}sica Te\'orica, Universidad Aut\'onoma de Madrid, 28049 Madrid, Spain}
    \affiliation{Instituto de F\'{\i}sica Te\'orica UAM-CSIC, Campus de Cantoblanco, 28049 Madrid, Spain}
\author{D.W.P.~Amaral} \affiliation{Department of Physics, Durham University, Durham DH1 3LE, UK}
\author{J.~Anczarski} \affiliation{SLAC National Accelerator Laboratory/Kavli Institute for Particle Astrophysics and Cosmology, Menlo Park, CA 94025, USA}
\author{T.~Aralis} \affiliation{SLAC National Accelerator Laboratory/Kavli Institute for Particle Astrophysics and Cosmology, Menlo Park, CA 94025, USA}
\author{T.~Aramaki} \affiliation{Department of Physics, Northeastern University, 360 Huntington Avenue, Boston, MA 02115, USA}
\author{I.~Ataee~Langroudy} \affiliation{Department of Physics and Astronomy, and the Mitchell Institute for Fundamental Physics and Astronomy, Texas A\&M University, College Station, TX 77843, USA}
\author{C.~Bathurst} \affiliation{Department of Physics, University of Florida, Gainesville, FL 32611, USA}
\author{R.~Bhattacharyya} \affiliation{Department of Physics and Astronomy, and the Mitchell Institute for Fundamental Physics and Astronomy, Texas A\&M University, College Station, TX 77843, USA}
% ...
\collaboration{SuperCDMS Collaboration}

\date{\today}

\begin{abstract}
Cryogenic calorimeters for low-mass dark matter searches have achieved sub-eV energy resolutions, driving advances in both low-energy calibration techniques and our understanding of detector physics. The energy deposition spectrum of gamma rays scattering off target materials exhibits step-like features, known as Compton steps, near the binding energies of atomic electrons. We demonstrate a successful use of Compton steps for sub-keV calibration of cryogenic silicon calorimeters, utilizing four SuperCDMS High-Voltage eV-resolution (HVeV) detectors operated with 0~V bias across the crystal. This new calibration at 0~V is compared with the established high-voltage calibration using optical photons. The comparison indicates that the detector response at 0~V is about 30\% weaker than expected, highlighting challenges in detector response modeling
for low-mass dark matter searches.

\end{abstract}
\maketitle

\section{Introduction}

The growing interest in theoretically motivated low-mass (sub-GeV/$c^2$) dark matter searches~\cite{albakry2022strategy,essig2022snowmass2021_1,essig2022snowmass2021_2} has driven the development of calorimeters with eV-scale energy resolutions~\cite{anthony2023applying,PhysRevLett.119.131802,HONG2020163757}. A promising design is the SuperCDMS High-Voltage eV-resolution (HVeV) cryogenic calorimeter, which employs Quasiparticle-trap-assisted Electrothermal-feedback Transition-edge sensors (QETs) deposited on semiconductor targets~\cite{HVeV_characterization}. The response of HVeVs to energy depositions is complex, necessitating data-driven calibration methods to measure their response to known energy depositions.

The SuperCDMS collaboration has previously calibrated HVeVs using electron recoils induced by optical photons while applying a high-voltage bias across the detector (the \textit{HV mode})~\cite{HVeV_characterization,HVeVR1,HVeVR2,Low_energy_events_HVeV,HVeVR3}. In the HV mode, the detector response exhibits distinct peaks corresponding to the number of electron-hole (e${}^-$h${}^+$) pairs, with each peak approximately representing the sum of the initial energy deposition and the energy supplied to the e$^-$h$^+$ pairs by the applied electric field.

The detector calibration with optical photons presents several challenges. First, the photon source requires a line-of-sight to the detector, posing a risk of radioactive background contamination in rare-event searches. Second, the detector response to surface energy depositions by optical photons, occurring on a micrometer scale, may differ from the response to bulk energy depositions. Finally, operating the photon source can introduce interferences with the detector readout, such as electronic cross-talk, which can be difficult to isolate.

This work presents a new calibration technique for cryogenic silicon calorimeters using Compton steps. These steps emerge in the differential cross section of Compton scattering at the binding energies of electrons. The absence of a final quantum state for the scattering process forbids the electrons with binding energies greater than the transferred energy to contribute to the cross section. The binding energies of electrons in silicon are approximately 100~eV for the L3 and L2 shells, 150~eV for the L1 shell, and 1.84~keV for the K shell~\cite{norcini2022precision}. These energies are in the range of interest for the SuperCDMS SNOLAB dark matter searches~\cite{albakry2022strategy}.

We calibrated four HVeV detectors using Compton steps while operating them at 0~V bias across the crystal. The 0~V calibration was compared with the optical photon calibration at 150~V bias. The response of detectors in the 0 V mode was observed to be approximately 30\% weaker than that of the HV operation for the same expected amount of phonon energy, a trend similar to previous observations~\cite{HVeV_characterization}.

This manuscript is organized as follows: Section \MakeUppercase{\romannumeral 2} reviews the experimental setup and data collection. Section \MakeUppercase{\romannumeral 3} details the simulations that provided the expected spectrum of energy depositions in the detectors. Sections \MakeUppercase{\romannumeral 4} and \MakeUppercase{\romannumeral 5} cover event reconstruction and selection, respectively. Section \MakeUppercase{\romannumeral 6} discusses the new calibration using the Compton steps at 0~V detector bias. Section \MakeUppercase{\romannumeral 7} covers the HV LED calibration. Section \MakeUppercase{\romannumeral 8} compares the two calibrations, and Section \MakeUppercase{\romannumeral 9} provides our conclusions.

\section{\label{sec:experiment_setup}Experimental setup and data collection}
The experiment was conducted at the Northwestern EXperimental Underground Site (NEXUS) facility at Fermilab under a rock overburden that is equivalent to 225 meters of water~\cite{adamson2015observation}. A dilution refrigerator maintained the HVeV detectors at a stable temperature of 11~mK. The refrigerator was shielded with lead on the sides and bottom, with partial coverage on top.

This study utilized four NEXUS-Fermilab (NF) HVeV detectors, each consisting of a 0.93~gram high-purity silicon substrate measuring 1.0 × 1.0 × 0.4 cm${}^3$. One face of each substrate was covered by tungsten and aluminum QETs, while the opposite face had an aluminum grid covering approximately 5\% of the area. The QET side of each detector was grounded to the refrigerator chassis, and the aluminum grid could be voltage-biased relative to the chassis. The QETs on the HVeVs were arranged into two concentric square channels of equal surface area, with the QETs within each channel connected in parallel. The detectors in this study varied in the total surface area covered by QETs, with the naming NF-E, NF-H, and NF-C corresponding to increasing surface coverage values of approximately 8.5~mm$^2$, 21.6~mm$^2$, and 31.8~mm$^2$, respectively. The QET signals were read out using Superconducting Quantum Interference Device (SQUID)-based amplifiers, with the output measured in microamperes. This study employed one NF-E, one NF-H, and two NF-C detectors.

\begin{figure}
\includegraphics[width=0.49\textwidth]{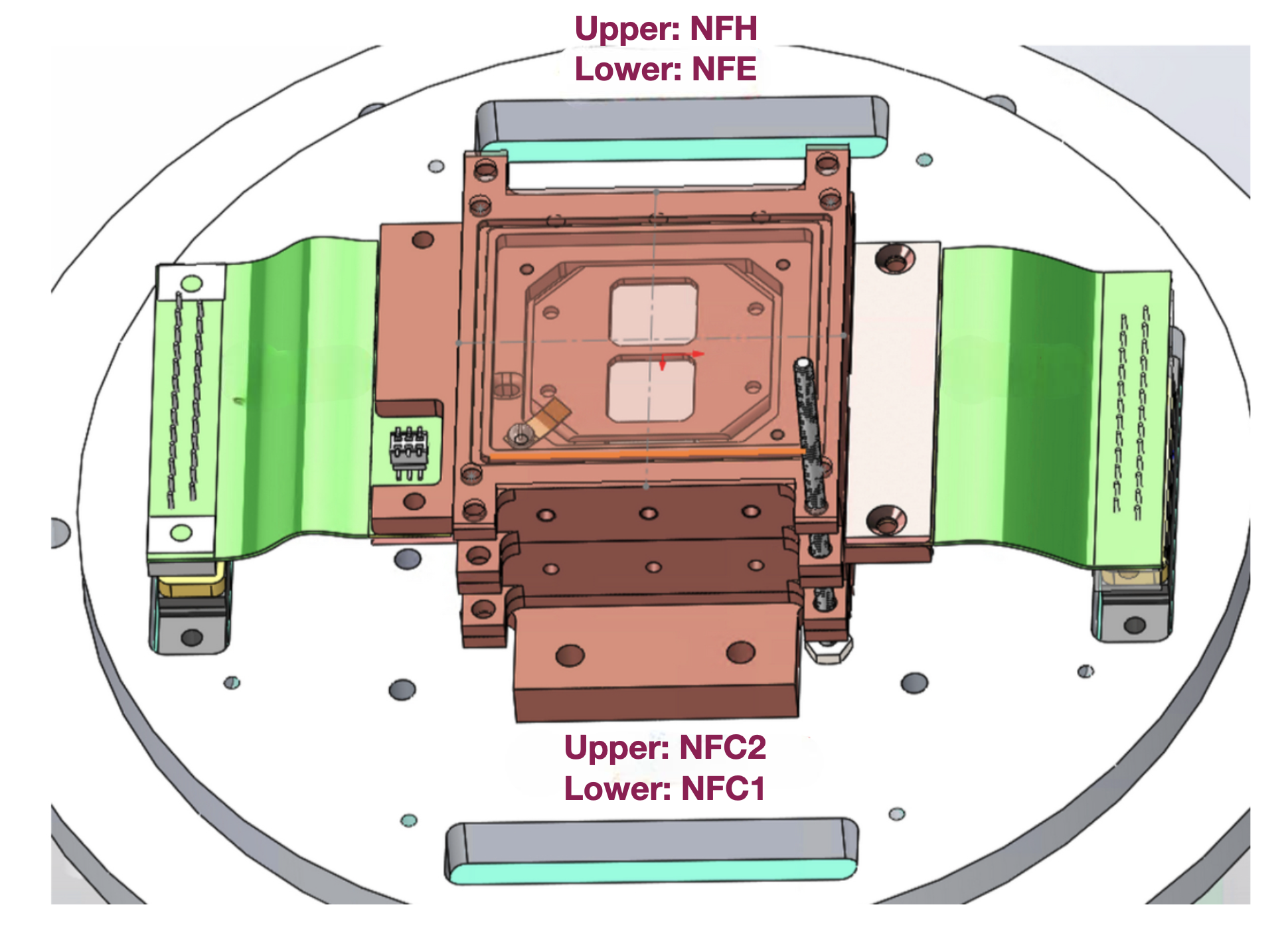}
\caption{Schematic of the detector tower configuration during data acquisition without the LED modules. The top lid is not drawn. The top detector box contains the NF-H and NF-C2 detectors, while the bottom box houses the NF-E and NF-C1 detectors, which are not visible in the schematic.}
\label{fig:NEXUS_R13_tower}
\end{figure}

The detectors were installed inside a copper housing that was thermally coupled to the mixing chamber of the dilution refrigerator. The housing consisted of two copper boxes, each with two detector slots, stacked on top of each other. The upper box contained detectors NF-H and NF-C2, while the lower box housed NF-E and NF-C1, as illustrated in Fig.~\ref{fig:NEXUS_R13_tower}. To facilitate the study of coincident energy depositions between the detectors, the QET sides of the detectors in the upper and lower boxes faced each other through an opening. The detector housing was designed with minimal printed circuit board (PCB) components, driven by evidence that this material causes excess low-energy background~\cite{Low_energy_events_HVeV}. The experiment was conducted in two configurations. The Compton calibration setup used the standard detector housing described above. In the LED calibration setup, two additional levels—one above and one below the detector boxes—were added to accommodate LED modules for calibration with approximately 2~eV optical photons. Each detector was illuminated by a dedicated LED through a pinhole aiming at the center of the aluminum grid side. Pinholes were covered with an infrared filter (SCHOTT KG3) to eliminate thermal radiation from the activated warm LED.

Four types of data were collected: (a) 2 days of calibration data with an external $3~\mathrm{MBq}$ $^{137}\mathrm{Cs}$ gamma source irradiating the detectors at a $0~\mathrm{V}$ bias across the substrate, (b) 4 days of background data without a radioactive source at a $0~\mathrm{V}$ bias, (c) 6 days of background data at a $100~\mathrm{V}$ bias, and (d) multiple sub-datasets with LEDs periodically flashing the detectors at a 150~V bias. Datasets (a), (b), and (c) were collected using the Compton calibration setup, while (d) was acquired using the LED calibration setup.

The gamma source was placed outside the refrigerator, at approximately the same height as the detectors and at a radial distance of $65~\mathrm{cm}$ from them. The LEDs were driven by a sinusoidal pulse with a half-period of approximately 1.7~microseconds, at a burst frequency of 10~Hz. The LEDs emitted photons with a mean energy of 2~eV, measured by a Thorlabs CCS100 spectrometer, with an intensity that varied across sub-datasets. The highest-intensity datasets were used to calibrate the detectors in the keV energy range. The low-intensity datasets were used for two studies: estimating the magnitude of the cross-talk between the detector readout cables and LED power lines, and investigating the effect of surface energy deposition by optical photons~\cite{improved_CTII_model}.

The NF-E detector, with the smallest Transition Edge Sensor (TES) volume, exhibited the strongest nonlinear energy response at higher energy depositions~\cite{HVeV_characterization}. The NF-E data was solely studied at low energies near the Compton L-steps, and no LED data was collected for this detector.

\section{Simulations}
The detector calibration with Compton steps was performed by comparing the experimental data with simulations. The energy depositions in the silicon detectors by 661.7 keV gamma rays emitted from the $^{137}$Cs source were simulated using \textsc{Geant4} (v10.07.p04)~\cite{GEANT4:2002zbu}, modeling the full experimental setup, including detector housings, the dilution refrigerator, and shielding layers. The Monash physics library was used to model Compton scattering~\cite{f8bba843a6e44af48b77e2489f067559,PhysRev.33.643}. We note that the Monash library uses the Relativistic Impulse Approximation~\cite{PhysRevA.2.415,PhysRevB.12.3136}, which has been shown to deviate from experimental data at energies below 500~eV~\cite{norcini2022precision}. The Geant4 simulation predicts that, in the 50 eV to 2 keV energy range, Compton scattering from the external calibration source was the primary contributor to energy depositions in the detectors. At energies below 500~eV, the spectrum of energy depositions from the Geant4 simulation was replaced with \textit{ab initio} calculations of the Compton differential cross section using \textsc{FEFF} (v10)~\cite{kas2021advanced,REHR2009548,RevModPhys.72.621,feff4,feff5}. The \textsc{FEFF} package computes electronic excitation probabilities for an atom within a cluster, which can be used to calculate the Compton differential cross section~\cite{norcini2022precision}.

\begin{figure}
\includegraphics[width=\linewidth]{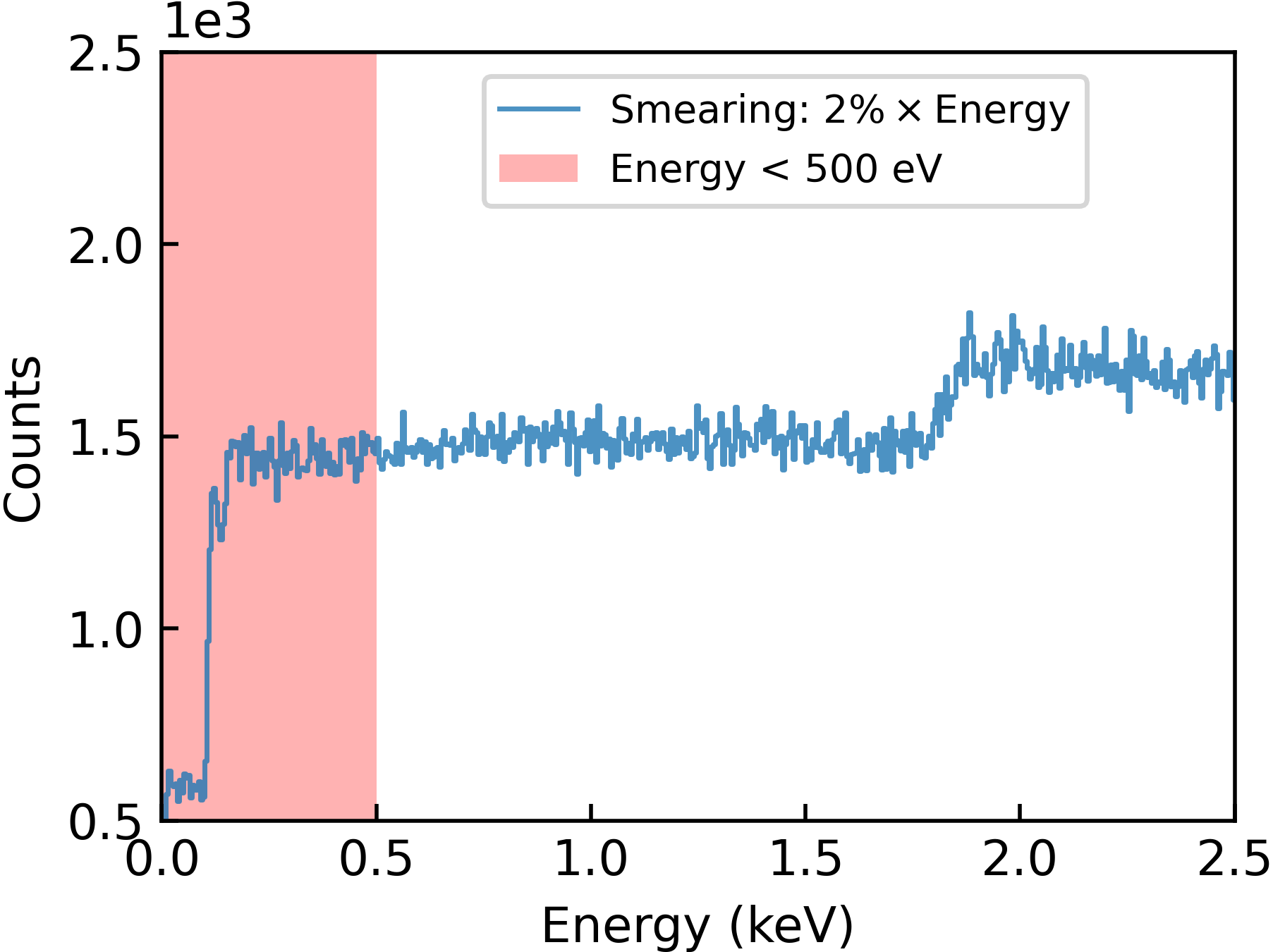}
\caption{The spectrum of energy depositions in HVeV detectors based on the Geant4 simulation. A Gaussian smearing with an energy-dependent width has been applied to the spectrum. The smearing corresponds to 2\% of the deposited energy, approximating the detector performance that is observed in previous studies \cite{HVeV_characterization}.}
\label{fig:geant4_high_energy}
\end{figure}

The detectors were positioned close to each other inside the housing, with no passive material in between. The spectra of energy deposition from the Geant4 simulation were compared between detectors using the Kolmogorov–Smirnov (KS) test~\cite{16e7f618-c06b-3d10-8705-1086b218d827}, and were found to be consistent with $p$-values greater than 10\%. The average of the Geant4 spectra across detectors, as shown in Fig.~\ref{fig:geant4_high_energy}, was used to model the data above 500 eV for all detectors.

We calculate the Compton scattering differential cross section between 50~eV and 500~eV using
\begin{equation}
\label{differential_xsection}
\frac{d\sigma}{dw} = r_0^2\int d\Omega  \left(\frac{1+\cos^2\theta}{2}\right)\left(1-\frac{w}{w_i}\right)\sum_{nl} S_{nl}(q,w),
\end{equation}
where $r_0$ is the classical electron radius, $\Omega$ is the scattering solid angle, $\theta$ is the scattering angle, $w$ and $q$ are the transferred energy and momentum, $w_i$ is the incident photon energy, and $S_{nl}(q, w)$ is the dynamic structure factor (DSF) with indices $n$ and $l$ indicating the quantum states of the target electrons. 

The FEFF calculation provided the DSF for the electronic shells, characterizing the relation between the permitted energy and the momentum transfer values based on the initial quantum state of the target electron. The FEFF calculation was performed with a silicon crystal consisting of 71~atoms, generated using the Web\-Atoms tool~\cite{Ravel:hf5152}. The calculation was repeated with a crystal consisting of 35~atoms, identical to the one employed in Ref.~\cite{norcini2022precision}, and obtained consistent results. We calculated the DSF in the energy range of interest for each of the three L-subshells\footnote{For each electronic shell, the XANES option was selected at energies up to 40~eV above the Compton step, and the EXAFS option was selected at higher energies \cite{kas2021advanced,norcini2022precision}.}.

\begin{figure}
\includegraphics[width=\linewidth]{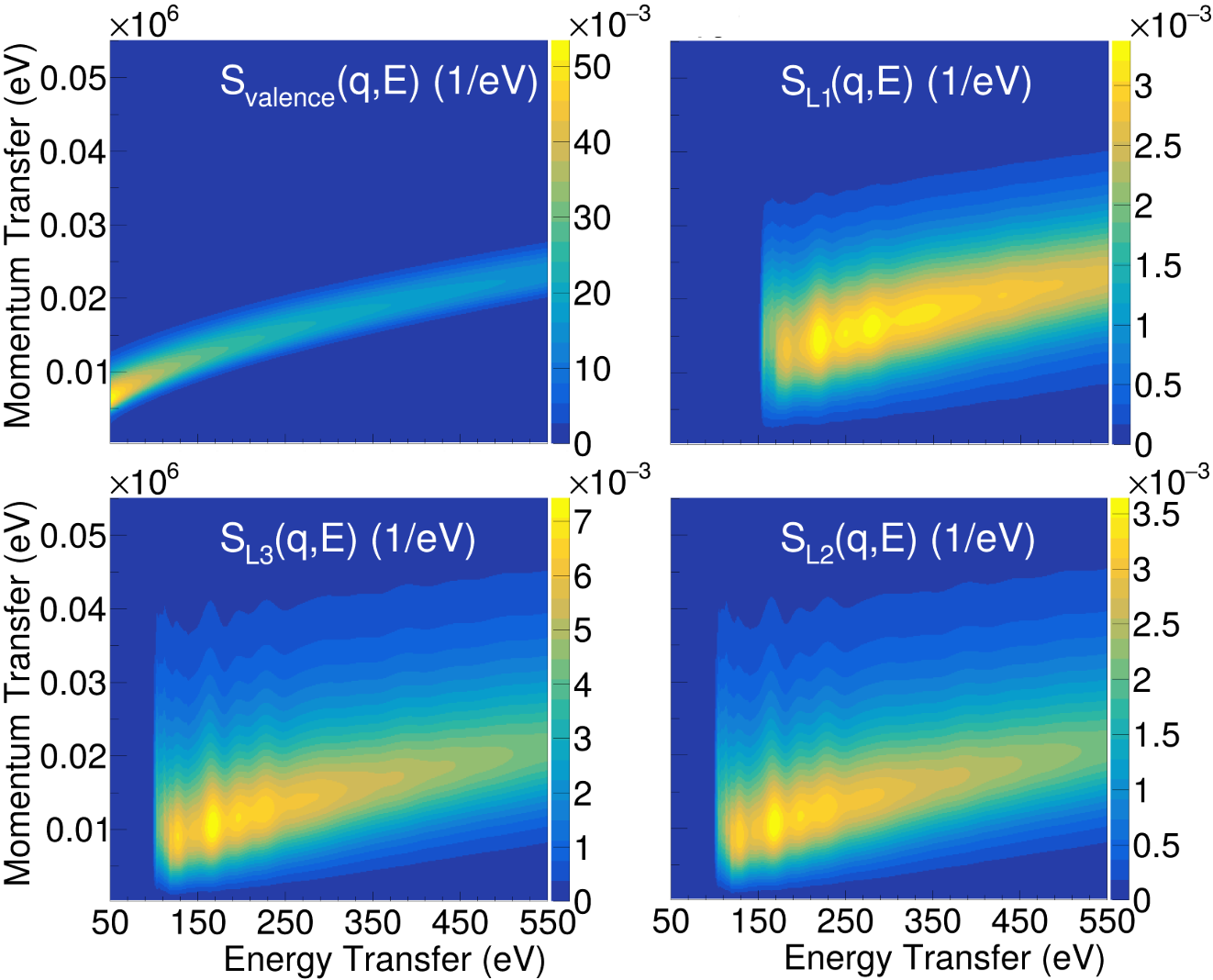}
\caption{Dynamic structure factors (DSF) computed using FEFF calculations for electrons in various shells. The energy and momentum transfer values are shown on the $\it{x}$ and $\it{y}$ axes, respectively. The color shows the magnitude of the DSF~\cite{cp_correction}.}
\label{fig:dsf}
\end{figure}

The contribution of the valence electrons to the cross section was considered by calculating their Compton profile using FEFF~\cite{PhysRevB.85.115135,norcini2022precision}. The Compton profile represents the density of electrons as a function of their projected momentum along the direction of momentum transfer. The calculated Compton profile was scaled with a constant multiplicative factor to obtain the correct number of valence electrons per unit cell in silicon, after which the corresponding DSF was computed~\cite{cp_correction}. The DSFs for the valence and L-shell electrons are shown in Fig.~\ref{fig:dsf}.

The integration in Eqn.~\ref{differential_xsection} was performed by replacing the scattering angle dependence with the momentum transfer, utilizing the energy and momentum conservation equations. In the energy range of gamma rays reaching the detectors, the shape of the Compton differential cross section remained consistent with that of the highest-energy gamma rays. Consequently, $w_i$ was set to 661.7~keV, corresponding to the primary energy of gamma rays from the $^{137}$Cs source.

\begin{figure}
\includegraphics[width=\linewidth]{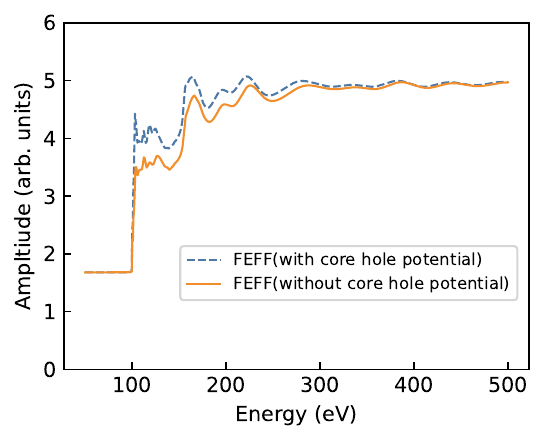}
\caption{Differential cross sections for Compton scattering between 50 eV and 500 eV, computed using Eqn.~\ref{differential_xsection} and DSFs from FEFF. The blue curve includes the effect of the vacated core-hole potential, while the orange curve excludes it.}
\label{fig:low-energy_sim} 
\end{figure}

The FEFF calculations of the DSF can either include or exclude the effect of the vacated core hole potential on the scattered electron. There is limited evidence showing that the FEFF calculations better match data without modeling the core-hole potential in silicon, including a measurement using skipper-CCDs~\cite{PhysRevB.75.075118,Sternemann:fh5381}. The DSF calculation without the core hole potential modeling was used to develop the primary Compton scattering differential cross section. The calculation with the core hole potential was used to quantify the impact of FEFF assumptions on the Compton step calibration. The differential cross section functions computed between 50~eV and 500~eV using Eqn.~\ref{differential_xsection} are shown in Fig.~\ref{fig:low-energy_sim}.

\section{Event reconstruction}
\label{sec:event reconstruction}
The experimental data were continuously saved to disk in segments of 0.5~seconds, referred to as traces, at a sampling frequency of 156.25~kHz. The readouts from the inner and outer channels of each HVeV detector were summed and passed through the offline triggering and processing algorithms. A threshold trigger was applied to the summed traces of each detector after filtering them with a kernel that took the form of the first derivative of a Gaussian function with a width of 38.4~microseconds. The trigger point for each pulse was set at the maximum of the filtered trace, and a 13.1-millisecond time window centered on the trigger point represented each event.

The constant firing frequency of the LED (10~Hz) and the timestamps of threshold-triggered LED events were used to identify the timing of all LED events. A custom logic ensured full triggering efficiency of LED events independent of the trigger threshold.

The optimal filter (OF) algorithm in the frequency domain was used to determine the amplitudes of triggered pulses, with the amplitudes serving as the energy estimator~\cite{OF_reference}. The OF algorithm can include a free time-shift parameter to correct for any misalignment between the pulse and the template~\cite{kurinsky_thesis}. We use $A_\text{OF}$ and $A_\text{OF0}$ to denote the pulse amplitude estimators with and without using the time-shift parameter. The pulse template of the OF algorithm was constructed by averaging events with a total phonon energy of approximately 100~eV from a 100~V dataset \footnote{Pulse shapes were found to be consistent between the 0~V and high-voltage (HV) datasets.}. The noise power spectral density (PSD) was obtained using randomly triggered events with no pulse-like features from the 100~V dataset. The random events with pulse-like features were rejected based on their mean, standard deviation, slope, and skewness.

The processing provided the pulse amplitudes $A_\text{OF}$, $A_\text{OF0}$, and $\chi^2$ per degree of freedom values ($\chi^2$/NDF). The $\chi^2$/NDF values quantified the similarity of the pulses with the processing template. We calibrated the \(A_{\text{OF}}\) values as the primary energy estimator.

\section{Data selection}
We defined data selection criteria tailored to the L-step and K-step searches. The L-step search window was selected to be within an energy range of 50~eV and 300~eV, based on an approximate calibration of the 0~V data. The approximate calibration was derived using the pulse amplitude values of events at approximately 100~eV~\cite{HVeVR3} in 100~V data. The K-step search was independent of the approximate calibration, as its signature was distinguishable in the uncalibrated data. A summary of the data selection criteria is provided in Table~\ref{table:list_cuts}.

\begin{table}
\caption{Summary of data selection criteria. Under the \textbf{Type} column, Livetime selections remove time intervals from the data independent of individual events, and Event selections eliminate individual events. The \textbf{Domain} column shows whether the selection is informed by the data processing setup (Processing), determined separately for short sub-segments spanning a few hours (Series-based), or defined after combining all datasets (All data).}
\centering
\begin{ruledtabular}
\begin{tabular}{lcc|c|c|}

\textbf{Selection name} & \textbf{Type} & \textbf{Domain} & \textbf{L-steps}   & \textbf{K-step} \\ 
\colrule
\addlinespace[0.5ex] 

Trigger deadtime & Livetime & Processing & \multicolumn{2}{c|}{\checkmark} \\ 
\hline
High-trigger rate & Livetime & Series-based & \multicolumn{2}{c|}{\checkmark} \\ 
\hline
Coincidence & Event & All data & \checkmark & \ding{55} \\ 
\hline
$\Delta$Time & Event & Processing & \multicolumn{2}{c|}{\checkmark} \\ 
\hline
Baseline & Event & Series-based & \multicolumn{2}{c|}{\checkmark} \\ 
\hline
OF-$\chi^2$ & Event & All data & \checkmark & \ding{55} \\ 
\hline
Pulse width & Event & All data & \ding{55} & \checkmark \\ 
\end{tabular}
\end{ruledtabular}
\label{table:list_cuts}
\end{table}

The \textbf{Trigger deadtime} selection removed 13.1~milliseconds (2050~samples) from the livetime at both ends of a trace (0.5~second data segments), equivalent to the full size of the pulse reconstruction window. This selection prevented processing artifacts at the edges of a trace, removing 5\% of the livetime.

Some datasets included time intervals with orders of magnitude greater trigger rates, originating from a train of square-shaped pulses. Devices from other experiments in the same dilution refrigerator produced radio-frequency noise causing these elevated-rate intervals. The \textbf{High-trigger rate} selection eliminated these intervals. We modeled the distribution of event counts per time interval with a Poisson function, where the interval size was 10~seconds for the calibration data and 100~seconds for the background data. The time intervals flagged for elevated trigger rates in any detector were discarded across all detectors. The \textbf{High-trigger rate} selection reduced the livetime of the calibration and background data by 6\% and 18\%, respectively.

A population of spike-like events was found between pairs of detectors when the energy deposition in one surpassed the keV range. We also observed spike-like events with distinct shapes compared to regular pulses, occurring simultaneously across all detectors. The \textbf{Coincidence} selection flagged any two events with trigger times within 96~$\mu$s (15 samples) across any of the four detectors. These non-signal-like events populated the L-step energy region. The \textbf{Coincidence} selection removed flagged events within the L-step window, excluding the keV range to prevent removing energy depositions paired with spike-like events.

The OF method cannot reliably find the energy estimator values for overlapping events. The $\boldsymbol{\Delta}$\textbf{Time} selection rejected any two events in the same detector with trigger times closer than 6.6~ms (1039~samples). This selection ensured that no two triggers fell within the processing window of a single event.

\begin{figure}
\includegraphics[width=\linewidth]{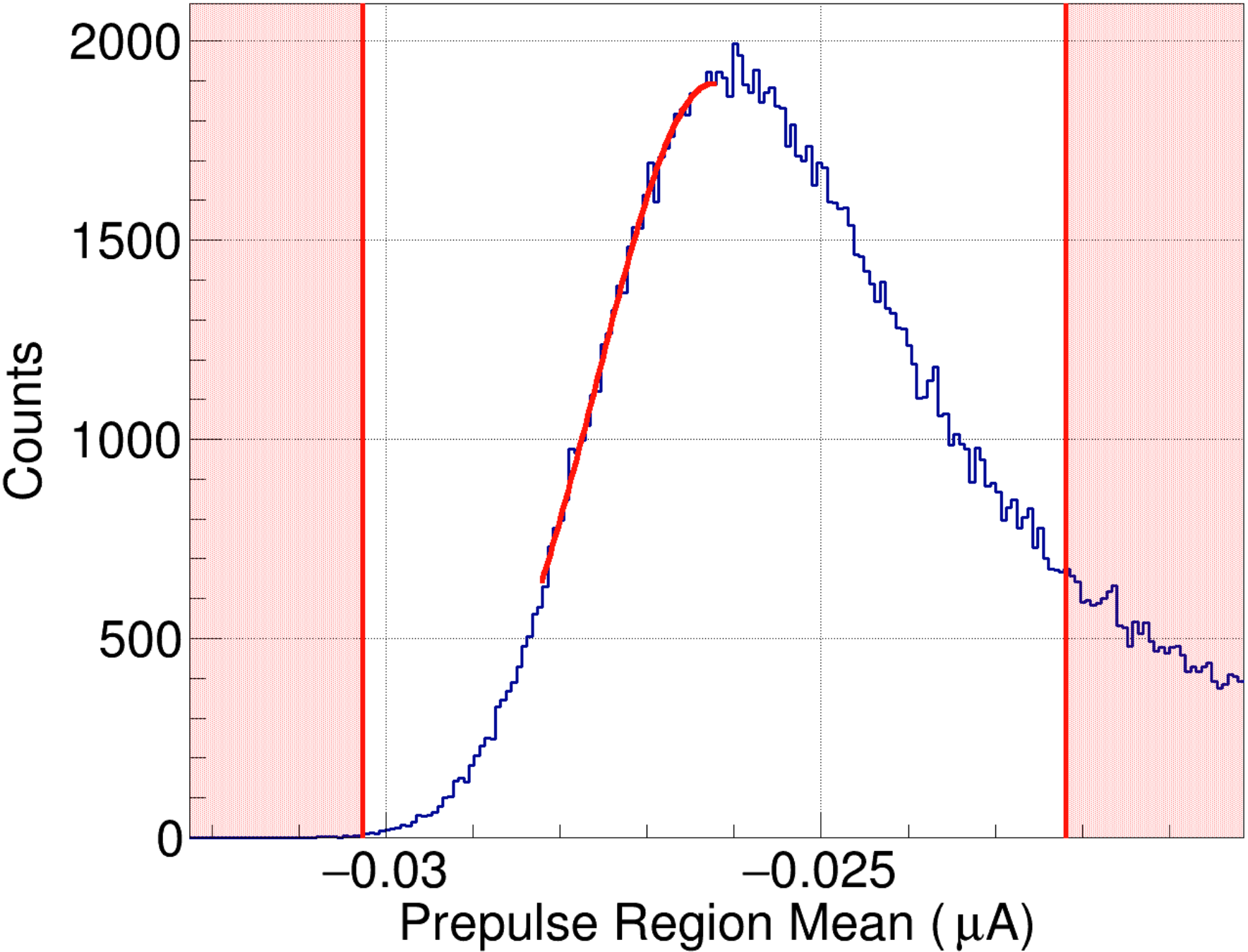}
\caption{The distribution of the average pre-pulse region values within the processing window for events in the calibration data. The fit of a Gaussian function to the left side of the distribution determines the mean value and standard deviation of the distribution. Events in the shaded regions were three standard deviations away from the mean of the Gaussian and were rejected.}
\label{fig:MeanBase}
\end{figure}

The calibration data contained many high-energy pulses with long tails, resulting in a high probability of events occurring before the detector had fully recovered from preceding events. The \textbf{Baseline} selection primarily removed events strongly affected by residual pulse tails from preceding events. The average values of traces in the pre-pulse region of events, denoted by $\nu_{\text{base}}$, were used to define this selection. In the absence of the pulse overlaps and under stable detector operation, the average values in the pre-pulse regions should follow a Gaussian distribution due to the trace noise. The presence of overlapping events distorted this Gaussian distribution, as shown for one dataset in Fig.~\ref{fig:MeanBase}. The \textbf{Baseline} selection modeled the left side of the distribution with a Gaussian function and rejected events with $\nu_{\text{base}}$ values exceeding three standard deviations from the Gaussian mean.

The skewness in Fig.~\ref{fig:MeanBase} indicated that the \textbf{Baseline} selection reduced but did not fully eliminate overlapping events. The calibration bias from residual overlapping events was studied using data collected at a 100~V detector bias and with the radioactive source in place. The 100~eV events from the first e$^{-}$h$^{+}$ peak were selected and filtered with the \textbf{Baseline} selection. The variation in pulse amplitude values of accepted events as a function of $\nu_{\text{base}}$ was found to introduce negligible uncertainty in the calibration compared to other sources.

\begin{figure}
\centering
\includegraphics[width=\linewidth]{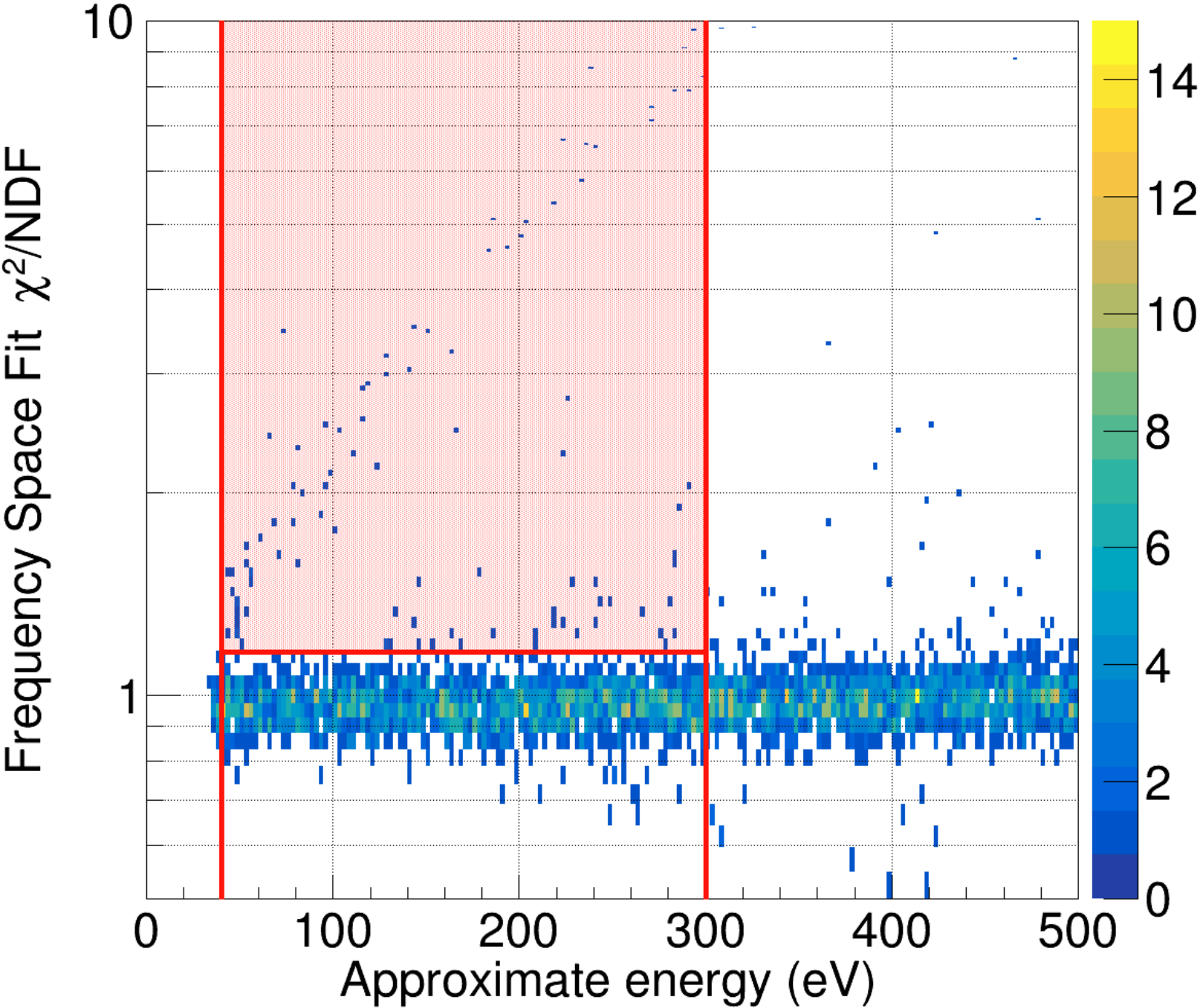}
\caption{Distribution of \(\chi^2\) per degree of freedom values as a function of approximate energy. Events between 40~eV and 300~eV were selected. The distribution of \(\chi^2/\text{NDF}\) values for the selected events was modeled with a Gaussian function to determine the mean and standard deviation. Events above three standard deviations from the Gaussian mean, as shown by the red shaded area, were rejected.}
\label{fig:chi2}
\end{figure}

A small population of events had elevated tails, resulting from direct energy depositions in QETs~\cite{HONG2020163757}. The \textbf{OF-}$\bm{\chi^2}$ selection eliminated the elevated-tail pulses. These events form a diagonal distribution within the L-step search window, when comparing the $\chi^2$/NDF values to the energy, as shown in Fig.~\ref{fig:chi2}. The \(\chi^2/\text{NDF}\) distribution for events in the L-step search window was modeled with a Gaussian function. The events with \(\chi^2/\text{NDF}\) values exceeding three standard deviations above the mean were rejected.

The detector response in the keV-scale energy range differed from that of the low-energy events used to develop the OF template. That led to a gradual increase in the mean and standard deviation of the \(\chi^2/\text{NDF}\) distribution as a function of energy. The pulse amplitude was still selected as an energy estimator; however, events with elevated tails were identified using the \textbf{pulse width} selection. The pulse width values were measured at 30\% of the maximum pulse height and anomalously wide pulses were rejected.

\begin{figure}
\centering
        \includegraphics[width=1\linewidth]{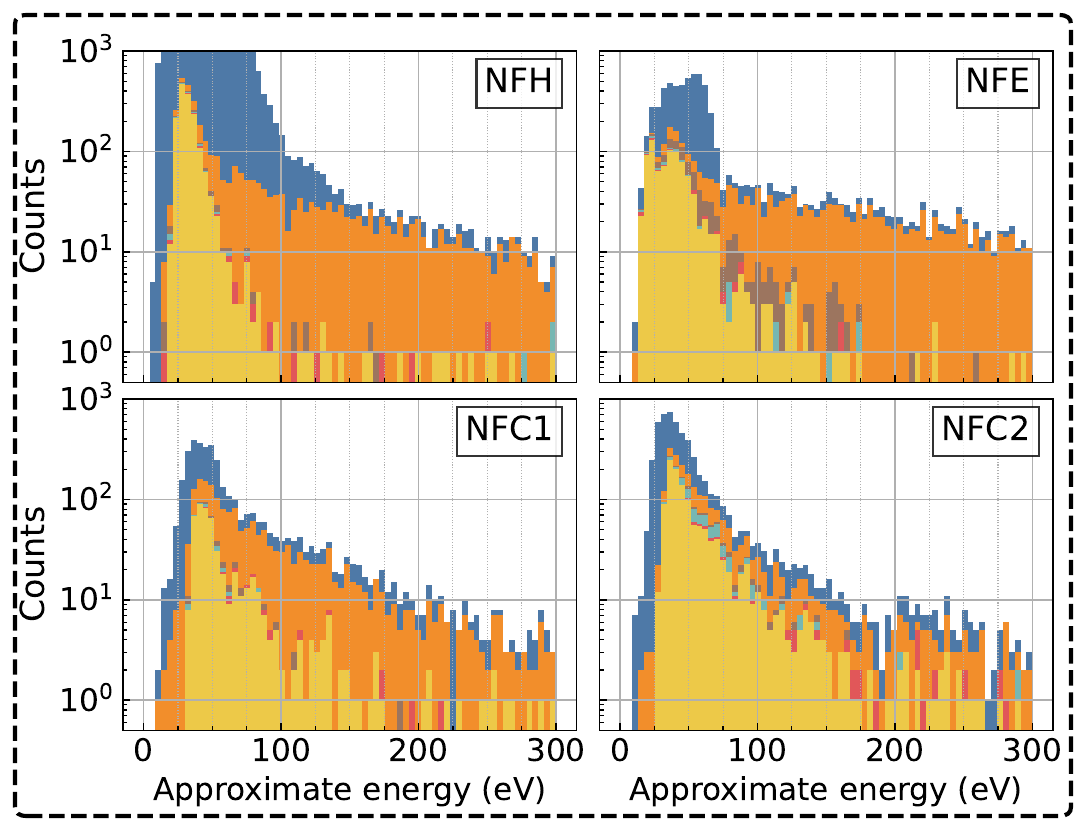}
    \includegraphics[width=1\linewidth]{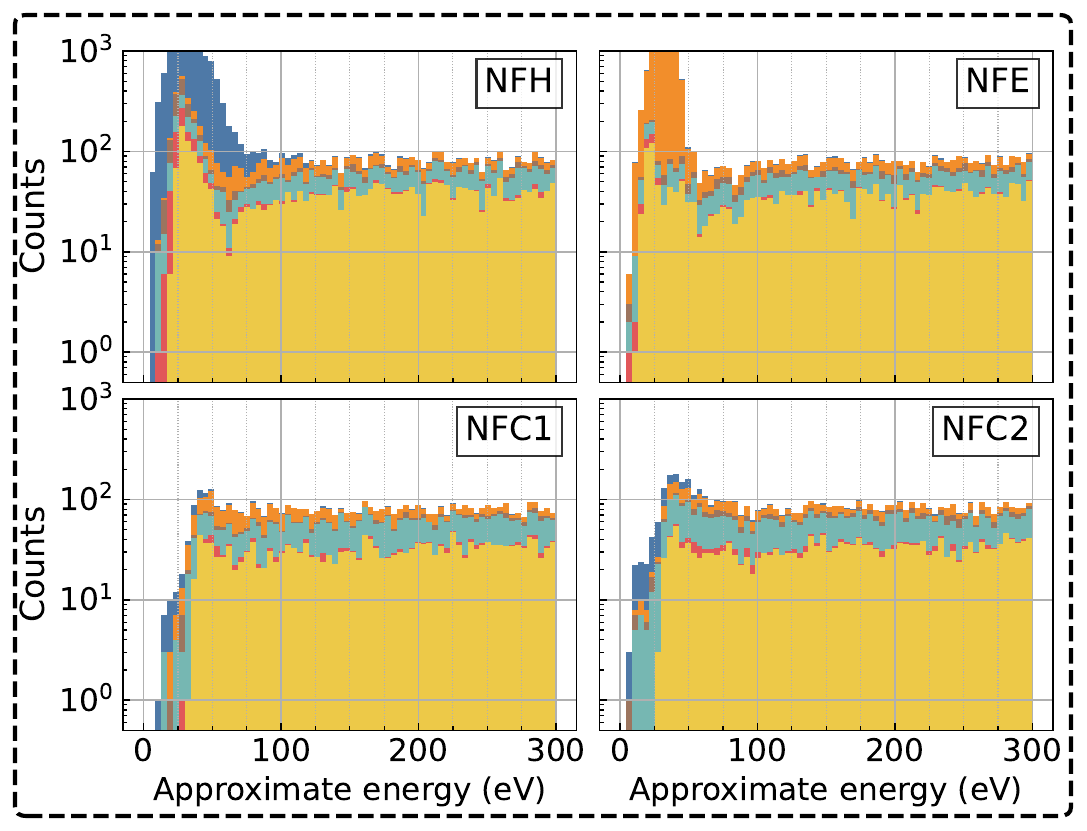}
    \includegraphics[width=1\linewidth]{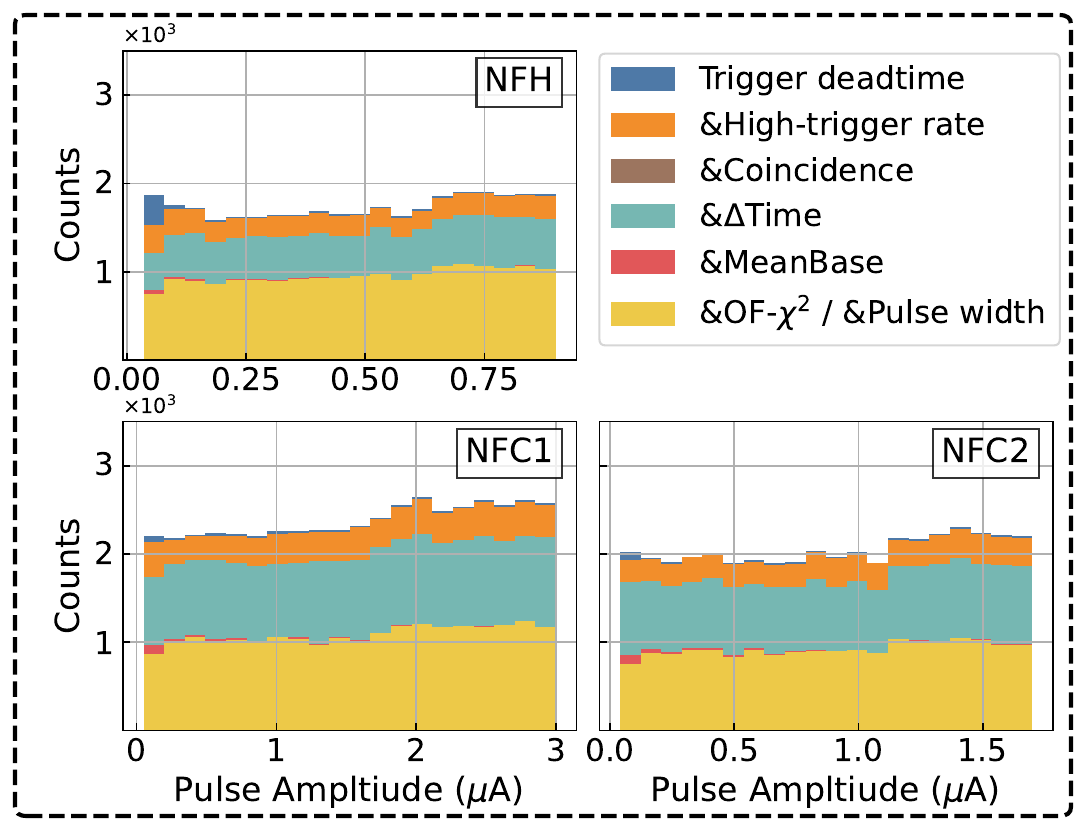}
  \caption{Spectra of passing events in 0~V datasets after sequentially applying the selections. Each color represents the passing events up to and including the selection that is named in the legend. Top: Low-energy background data ($\sim$4~d). Middle and Bottom: Low-energy and high-energy ranges of the calibration data ($\sim$2~d). The \textbf{Coincidence} selection is not applied in the high-energy range, and the \textbf{OF-}\bm{$\chi^2$} selection is replaced with the \textbf{Pulse width} selection, as indicated in the legend by `/ \&Pulse width'.}
\label{fig:high_energy_spectum}
\end{figure}

These selections removed poorly reconstructed events while maintaining an energy-independent signal efficiency within the K- and L-steps search windows separately. The livetime selections rejected time intervals rather than individual events, making it independent of reconstructed energy. The mean free path of 100~keV-scale gamma rays in silicon is on the order of a few centimeters, making it unlikely for low-energy Compton scatters to produce true coincidences across multiple detectors. Thus, the \textbf{Coincidence} selection in the low-energy region removed non-signal-like events without introducing an energy-dependent signal rejection. The $\boldsymbol{\Delta}$\textbf{Time} and \textbf{Baseline} selections rejected the largest fraction of data, primarily removing overlapping events. Using a Monte Carlo simulation with pseudo-pulses injected randomly into a data stream, we confirmed that rejecting overlapping events yields an energy-independent signal rejection. The \textbf{OF-}\bm{$\chi^2$} and \textbf{Pulse width} selections removed elevated-tail events with distinct shapes compared to signal-like events. The spectra of passing events by sequentially applying the data selections are shown in Fig.~\ref{fig:high_energy_spectum}.

We modeled the triggering efficiency of the detectors using 
\begin{equation} 
\epsilon( A_{\text{OF}} \mid p_1, p_2) = \frac{1}{1+\exp{\left(-\frac{A_{\text{OF}}-p_1}{p_2}\right)}},
\label{efficiency equ}
\end{equation}
where $p_1$ and $p_2$ were free parameters. The parameters of the triggering efficiency function were constrained using a Monte Carlo simulation. The simulation injected signal templates with varying amplitudes into randomly triggered events with no pulse-like features. The pseudo-events were processed and triggered, and the fraction of triggered pseudo-events was modeled using the triggering function. We used the constraints on the parameters of the triggering efficiency function in the search for the Compton L-steps in the data.

\section{0~V Compton steps calibration}
We measured the pulse amplitude values at the energies corresponding to the silicon Compton steps. The pulse amplitudes, along with their equivalent energy values, provided the data points used to find a calibration function up to 1.8~keV for each detector.

In the L-steps search window, the FEFF simulation was stretched and combined with a data-driven background model, described by an analytical function, to fit the calibration data. The analytical function was constrained by the background data. The stretch parameter provided a linear relation between pulse amplitudes and energies below 300~eV. This relation was used to read the pulse amplitude values at the position of the Compton L-steps. 

The Compton K-step was located in the Geant4 simulation and experimental data separately using an error function with an elevated baseline. The center of the error function in the simulation and experiment data provided the energy and its corresponding pulse amplitude values for each detector, respectively.

A detailed description of the procedure for locating the steps is provided below.

\subsection{Compton L-steps}
The background data collected without the radioactive source exhibited a rapidly falling spectrum, as shown in the top panel of Fig.~\ref{fig:high_energy_spectum}. A similar background spectrum has been observed by various cryogenic crystal calorimeters and is commonly referred to as the Low Energy Excess (LEE) \cite{angloher2023latest,PhysRevD.107.122003,PhysRevD.106.083009,10.1063/5.0222654,anthony2024low,PhysRevLett.127.061801,adari2022excess}.

To constrain the shape of the LEE spectrum, calibration data were collected between two intervals of background data over a period of 6 days. The spectra of the background data, collected before and after the calibration data, remained consistent as suggested by a KS test, which yielded a p-value greater than 20\%.

The shape of the LEE spectra was assumed to be consistent between the calibration and background data. Similarly, the Compton scattering spectrum from gamma rays—whether from the environment or the radioactive source—was assumed to have an identical shape. We modeled both the background and calibration data using a probability density function (PDF) of the form
\begin{equation}
    \label{eqn:pdf}
    \begin{split}
    & \text{PDF}_k(A_{\text{OF}} \mid C, \sigma, d, N^{\text{Comp}}_k,N^{\text{LEE}}_k) \,= \\ 
    & \quad \  \frac{N^{\text{Comp}}_k}{N_k}\cdot {\rm PDF}^{\rm Comp}(C\cdot A_{\rm OF} \mid \sigma) \\
    & \,+ \, \frac{N^{\text{LEE}}_k}{N_k} \cdot {\rm PDF}^{\rm LEE}(A_{\text{OF}}\mid d),
    \end{split}
\end{equation}
where index $k$ refers to either background or calibration data, PDF$^{\rm Comp}$ is the FEFF differential cross section normalized to have an area of unity, PDF$^{\rm LEE}$ is a normalized power-law function of the form $x^{-d}$ with $d$ as a free parameter, $N^{\text{Comp}}_k$ and $N^{\text{LEE}}_k$ are the numbers of events associated with the Compton scattering and the LEE spectra, and $N_k \, =\, N^{\text{Comp}}_k + N^{\text{LEE}}_k$ normalizes the overall PDF. The detector resolution was modeled using a Gaussian function with a constant width, $\sigma$, applied to the Compton differential cross section. In Eqn.~\ref{eqn:pdf}, $C$ is a linear calibration factor that relates the pulse amplitude values to deposited energies following
\begin{equation}
    E_{\rm OF} = C \cdot A_{\text{OF}}.
\end{equation}
 
\begin{figure}
\centering
    \includegraphics[width=\linewidth]{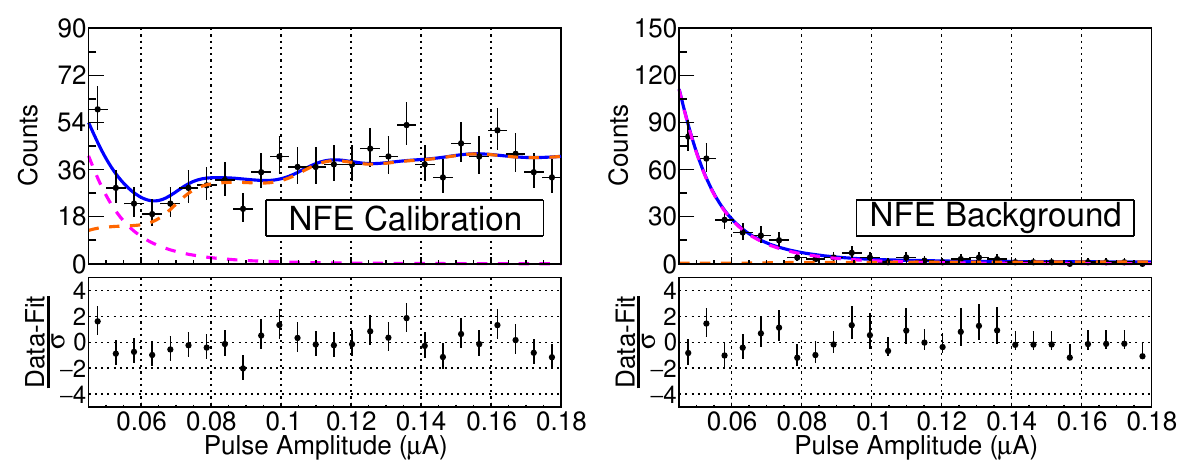}
    \includegraphics[width=\linewidth]{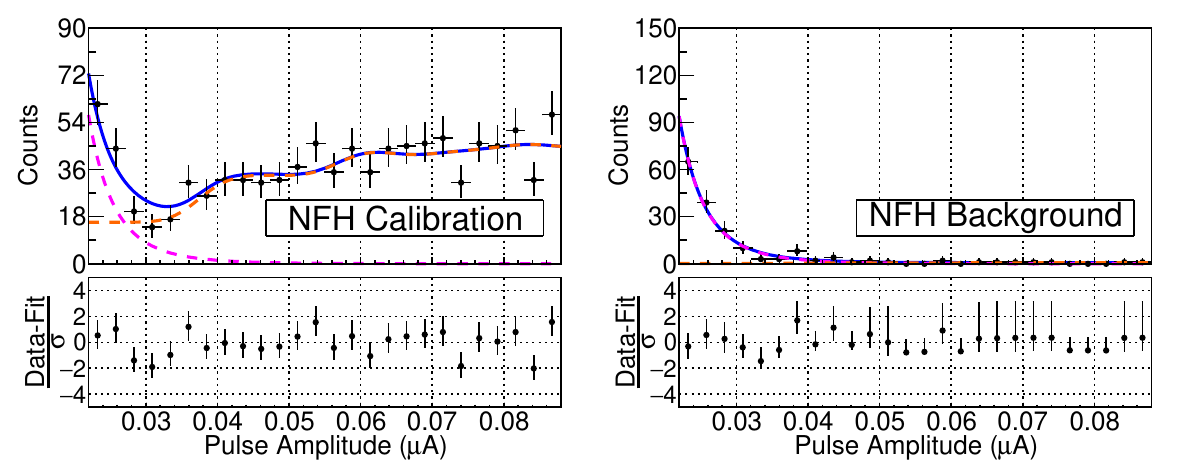}
    \includegraphics[width=\linewidth]{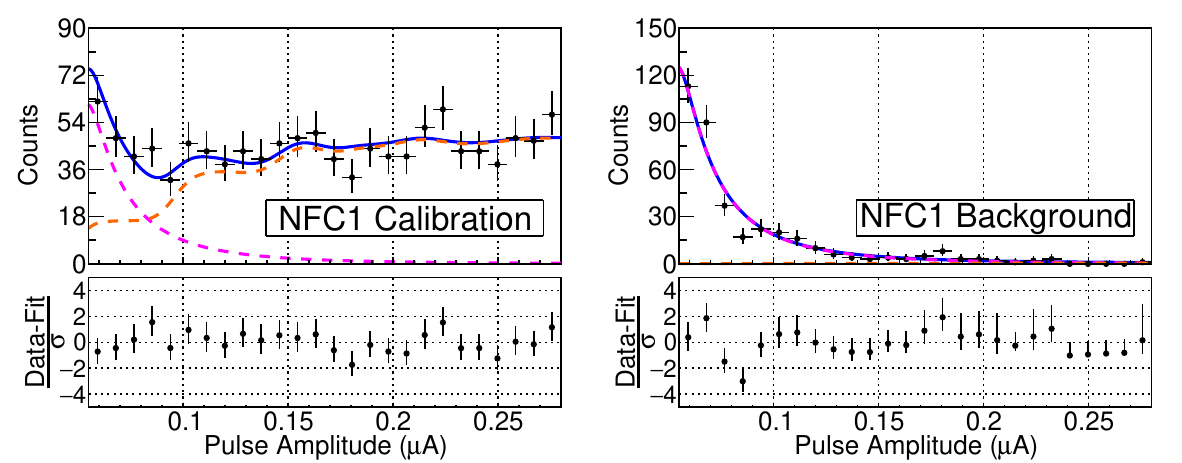}
    \includegraphics[width=\linewidth]{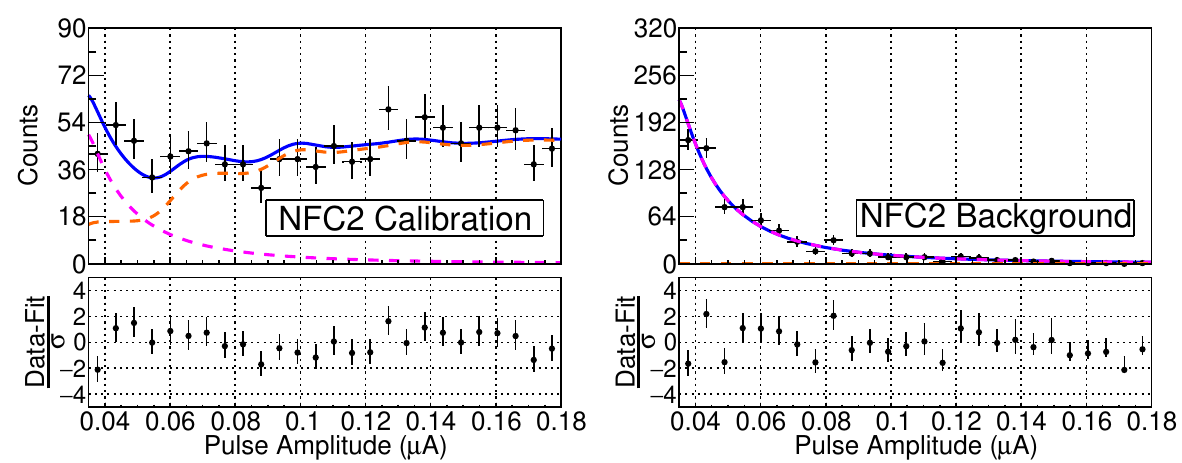}
  \caption{Simultaneous fit of background (right) and calibration (left) data for the silicon Compton L-steps. The solid blue lines represent the best fit, while the dashed orange and magenta lines indicate the contributions from Compton and LEE events, respectively. Datasets were modeled using the PDF in Eqn.~\ref{eqn:eff}. The parameters that control the shape of the PDFs were shared between the background and calibration data; however, the number of events were independent and free parameters.}
 \label{fig:compton_fits}
\end{figure}

The PDF in Eqn.~\ref{eqn:pdf} was adjusted using the triggering efficiency function in Eqn.~\ref{efficiency equ} following
\begin{equation}
    \label{eqn:eff}
    \text{PDF}^{\epsilon}_k(A_{\text{OF}}) = \frac{\text{PDF}_k(A_{\text{OF}}) \times {\epsilon}(A_{\text{OF}}\mid p_1, p_2)}{\int \text{PDF}_k(A_{\text{OF}}) \times {\epsilon}(A_{\text{OF}}\mid p_1, p_2) dA_{\text{OF}}},
\end{equation}
where the denominator normalizes the PDF in the L-step fit window. The likelihood function for the L-step fit is given by
\begin{equation}
\label{eqn:likelihood_trig_eff}
    \begin{split}
        L = &\mathcal{N}_2( \vec{\mu}, \boldsymbol{\Sigma} ; p_1, p_2) \\
            & \times \prod_{k}^{\rm bg, calib} {\rm Pois}(n_{k} \mid N_k(\Vec{\theta}_k)) \\
            & \times \ \ \prod_{i}^{n_{k}}  \ \ \text{PDF}^{\epsilon}_k(x_k^i \mid \Vec{\theta}_k,N^{\text{Comp}}_k,N^{\text{LEE}}_k),
    \end{split}
\end{equation}
where $\mathcal{N}_2$ is a two-dimensional Gaussian constraint with mean values $\vec{\mu}$ and a covariance matrix $\boldsymbol{\Sigma}$ that are taken from the standalone study of the triggering efficiency for parameters $p_1$ and $p_2$, $n_k$ is the number of observed events, $x^i_k$ represents the $A_{\text{OF}}$ values for events in the L-steps window, and $\Vec{\theta_k} = (C, \sigma, d, p_1, p_2)$ is a vector of free parameters that controls the shape of the PDFs. The data and the best fit curves are shown in Fig.~\ref{fig:compton_fits}. 

The likelihood in Eqn.~\ref{eqn:likelihood_trig_eff} was profiled to identify the systematic and statistical uncertainties of the calibration parameters~\cite{Cowan:1998ji}. Given that the asymmetries between the positive and negative 1$\sigma$ uncertainties were small, the larger of the two was used as an approximation for each detector. The calibration parameter was derived from fits using the no-core-hole FEFF PDF \cite{norcini2022precision}. We performed fits with the core-hole PDF and included half the difference between the resulting calibration factors as an additional systematic uncertainty. The total uncertainty was found by adding the profiled-likelihood and the FEFF uncertainties in quadrature, as summarized in Table~\ref{table:calibration_factors}.

\begin{table}
\caption{Summary of calibration factors from the L-step fits. The $A_{\text{OF}}$ values multiplied by the calibration factor ($C$) yield the deposited energies. The calibration is valid in the L-steps window up to 300~eV. The contributions to the uncertainty from the profile likelihood (PL) and the FEFF simulations are listed separately in columns 3 and 4, respectively, and combined in quadrature for the uncertainty in $C$.}
\begin{ruledtabular}
\begin{tabular}{lccc}
Detector & $C$ & PL (Stat \& Sys) & FEFF (Sys) \\
         & (keV/$\mu$A) & (keV/$\mu$A) & (keV/$\mu$A) \\
\colrule
\addlinespace[0.5ex] 
NF-E  & 1.5 $\pm$ 0.1 &  0.08  &  0.06 \\ 
NF-H  & 2.7 $\pm$ 0.2&  0.20 &  0.06  \\ 
NF-C1 & 1.06 $\pm$ 0.04&  0.04  &   0.01   \\ 
NF-C2 & 1.65 $\pm$ 0.08&  0.08   & 0.01  \\ 
\end{tabular}
\end{ruledtabular}
\label{table:calibration_factors}
\end{table}

\begin{figure*}
\centering
\includegraphics[width=0.329\linewidth]{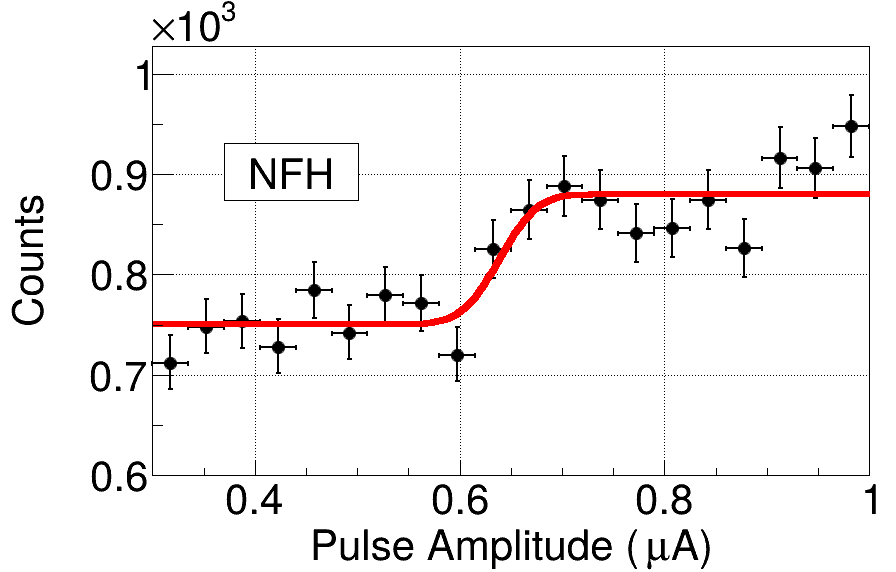}
\includegraphics[width=0.329\linewidth]{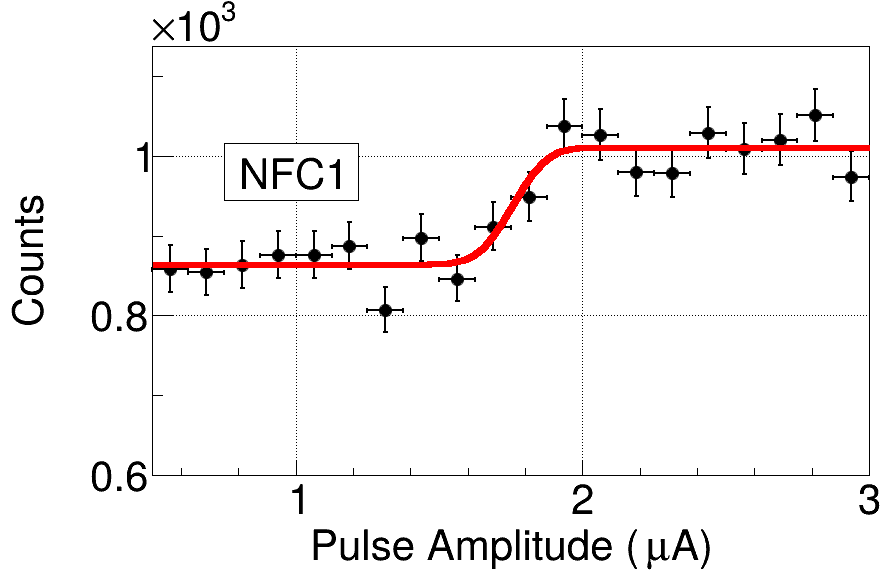}
\includegraphics[width=0.329\linewidth]{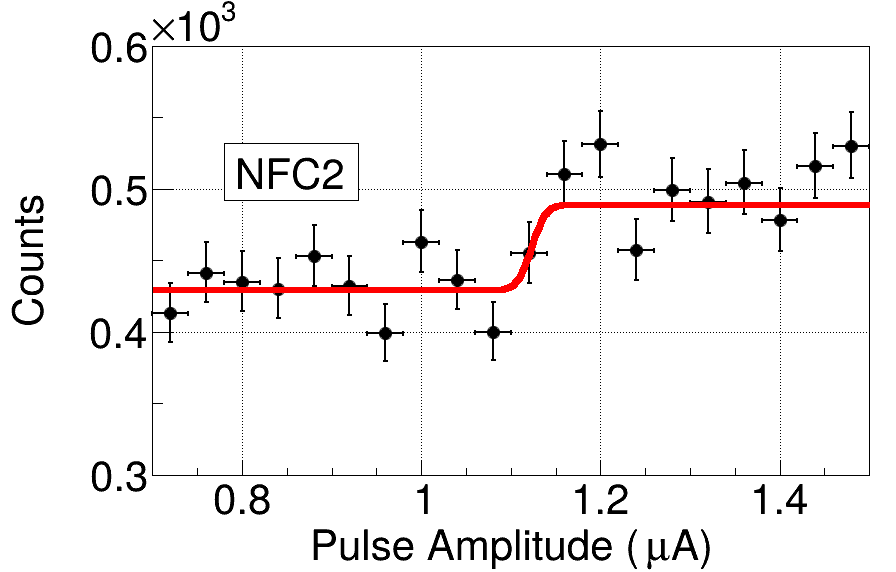}
\vspace{-0.5\baselineskip}
\caption{The spectra of $A_{\text{OF}}$ around the K-shell Compton steps for the NF-H, NF-C1, and NF-C2 detectors. The red curve is the fit to the data. The NF-E detector is excluded from the high-energy analysis due to its strong non-linear response at higher-energy depositions, attributed to its limited QET coverage.}
\label{0V Integral_spectrum}
\end{figure*}

\subsection{Compton K-step}
The Compton K-step in the data and Geant4 simulation were modeled separately from the L-steps using a modified error function that is given by
\begin{equation}
    f(a,b,\mu,\sigma)=\frac{a}{2}\cdot\left[1+\textrm{erf}\left(\frac{x-\mu}{\sqrt{2}\sigma}\right)\right]+b,
\end{equation}
where $a$ controls the height of the step, $b$ controls the baseline, $\mu$ controls the center of the step, and $\sigma$ controls the width, representing the smearing from the detector resolution. The function was normalized within a window around the K-step to represent a PDF. The center of the step in the data and simulation fits provided the pulse amplitude and its corresponding energy values. The modeling of the K-step in the calibration data is shown in Fig.~\ref{0V Integral_spectrum}. The resolution of the detectors at the location of the K-step was unknown prior to the fits. We adopted a nominal smearing value of 2\%, based on previous measurements performed with the NFC1 detector \cite{HVeV_characterization}. The Geant4 simulation was smeared with resolution values ranging from 1\% to 3\%, and the uncertainty of the K-step energy due to the smearing effect was found to be negligible compared to other sources.

\subsection{Calibration up to 2~keV}
We combined the information from the L- and K-step fits to obtain a calibration function up to 2~keV. The K-step fit provided a data point for a pulse amplitude value with a known energy. The L-steps fit was used to select another data point at 150~eV, serving as the second data point. The LED calibration that is discussed in Section~\ref{section: LED calibration} provided more data points and strongly preferred a quadratic calibration function for the response of the detectors. To maintain consistency, the Compton step data points were fitted by a quadratic function of the form 
\begin{equation}
    \label{eqn:non-linear}
    E_{\text{OF}}=\alpha A_{\text{OF}} + \beta A_{\text{OF}}^2.
\end{equation} 

The summary of measurements for $\alpha$ and $\beta$ is provided in Table~\ref{0V_calib_info}. The quadratic term in the 0~V calibration is statistically consistent with zero, indicating that the linear calibration function (used for the L-step analysis) and the quadratic calibration function yield consistent results within uncertainties. The 0~V and HV LED calibration curves will be presented and compared in Section~\ref{sec:compare}. Although a single representative data point was drawn from the L-steps fits to find the calibration up to 1.8~keV, the approximate locations of both Compton L-steps will be shown in Section~\ref{sec:compare}.

\begin{table}
\caption{Calibration factors to convert $A_{\text{OF}}$ values to energy for the 0~V datasets.}
\begin{ruledtabular}
\begin{tabular}{lll}
\textrm{Detector} & $\alpha$ (keV/$\mu$A)  & $\beta$ (keV/$(\mu \text{A})^2$) \\
\colrule
\addlinespace[0.5ex] 
NF-H   & $2.7 \pm 0.2$ &  $ \ \ 0.2 \pm 0.4$  \\

NF-C1  & $1.06 \pm 0.05$ & $-0.01  \pm 0.03$   \\

NF-C2  & $1.66 \pm 0.09$ & $ -0.03  \pm 0.08$  \\
\end{tabular}
\end{ruledtabular}
\label{0V_calib_info}
\end{table}

\section{High-voltage LED calibration} 
\label{section: LED calibration}
The detectors were operated at 150~V bias to take the LED calibration datasets~\cite{HVeVR1, HVeVR2, HVeVR3, HVeV_characterization}. The LED photons carried an average energy of approximately 2~eV, each generating a single e\(^-\)h\(^+\) pair once absorbed in silicon~\cite{eh_pair_creation_prob}. The voltage bias could drift the e\(^-\)h\(^+\) pairs across the crystal, amplifying the phonon signal through the Neganov-Trofimov-Luke (NTL) gain~\cite{neganov1978possibility,10.1063/1.341976}. 

\begin{figure}
\includegraphics[width=\linewidth]{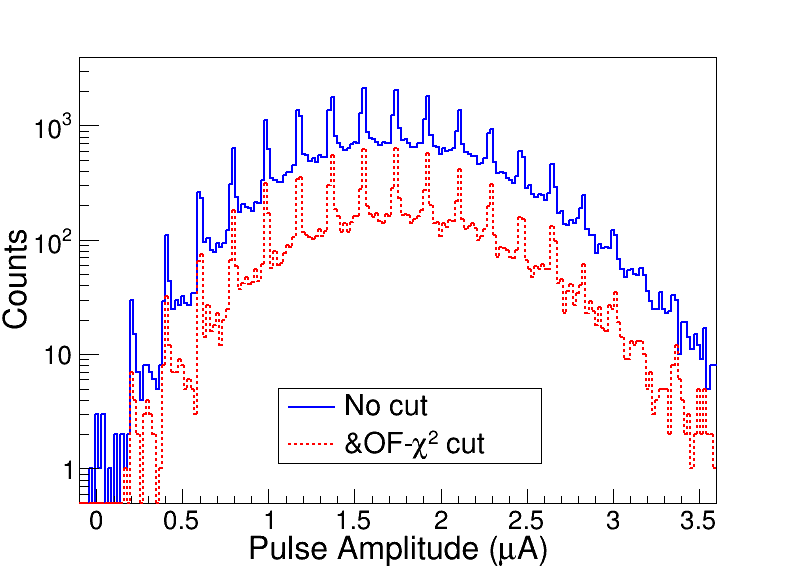}
\caption{Distribution of $A_{\text{OF}}$ for LED events before and after applying \bm{$\textbf{OF-}\chi^2$} based data selection for detector NF-C1.}
\label{fig:LED_spectrum}
\end{figure}

A quality selection based on the \textbf{OF}-\bm{$\chi^2$} was applied to the LED events. The spectrum of the events in one of the LED sub-datasets is shown in Fig.~\ref{fig:LED_spectrum}. The spectrum exhibits distinct peaks, with a distribution of events in between. The peaks resulted from the Poisson statistics for the number of photons reaching the detectors. Events occurring between the e$^-$h$^+$  peaks have previously been attributed to charge trapping and impact ionization processes~\cite{improved_CTII_model}. We use \(\lambda\) to denote the average number of photons per LED flash that reach a detector and undergo NTL amplification. The driving current of the LEDs was varied to change the intensity of flashes, and consequently $\lambda$, across various sub-datasets.

The pulse amplitude values at the center of individual e$^-$h$^+$ peaks served as the calibration data points. The center points were found by fitting each peak with a Gaussian function on top of a linear background. The energies at the center of e$^-$h$^+$ peaks were modeled by
\begin{equation}
\label{Eq: measured energy eh peak}
E_n = n_{\mathrm{eh}} \cdot (E_\text{photon} + e \cdot V_\text{bias}) + \bar{E}_\text{offset},
\end{equation}
where $E_\text{photon}$ and $e \cdot V_\text{bias}$ are the photon energy and the energy contributed by the NTL effect, respectively. The number of e$^-$h$^+$ pairs that contribute to NTL gain is given by $n_{\mathrm{eh}}$, and $\bar{E}_\text{offset}$ is an average residual energy associated with a fraction of e$^-$h$^+$ pairs that do not contribute to NTL gain. The surface trapping hypothesis has been previously used to explain $\bar{E}_\text{offset}$~\cite{improved_CTII_model}. This hypothesis suggests that a fraction of the e$^-$h$^+$ pairs recombine without traversing the voltage bias across the substrate, possibly due to random initial trajectories of charges relative to the applied electric field, as indicated by G4CMP simulations~\cite{geant4cmp,geant4cmp_code}.

The distribution of the number of surface trapped pairs comes from a convolution of the Poisson-distributed total number of produced pairs and a binomial extinction
process. It turns out the distribution of trapped pairs is also Poisson-distributed and
\emph{independent of $n_{\mathrm{eh}}$}~\cite{improved_CTII_model}. Since the variance of the energy offset is low compared to the energy resolution of our detectors, the overall effect is a uniform upward shift to the entire
spectrum by the average offset value, $\bar{E}_\text{offset}$.

\begin{table}[t]
\caption{
The energy offset in the detector response due to the surface trapping effect. The ratio \(\bar{E}_\text{offset}/\lambda\) is approximately 1~eV. }
\label{tab:surface_trap_offset}
\begin{ruledtabular}
\begin{tabular}{lcc}
\textrm{Detector} & \textrm{\(\lambda\)} & \textrm{$\bar{E}_\text{offset}$ (eV)} \\ 
\colrule
\addlinespace[0.5ex] 
NF-H   & 14.85 \(\pm\) 0.05 & 16 \(\pm\) 2  \\
NF-C1  & 9.66 \(\pm\) 0.05  & 10.6 \(\pm\) 0.7  \\
NF-C2  & 13.67 \(\pm\) 0.07 & 13 \(\pm\) 4  \\
\end{tabular}
\end{ruledtabular}
\end{table}

\begin{figure}[h!]
    \centering
    \includegraphics[width=\linewidth]{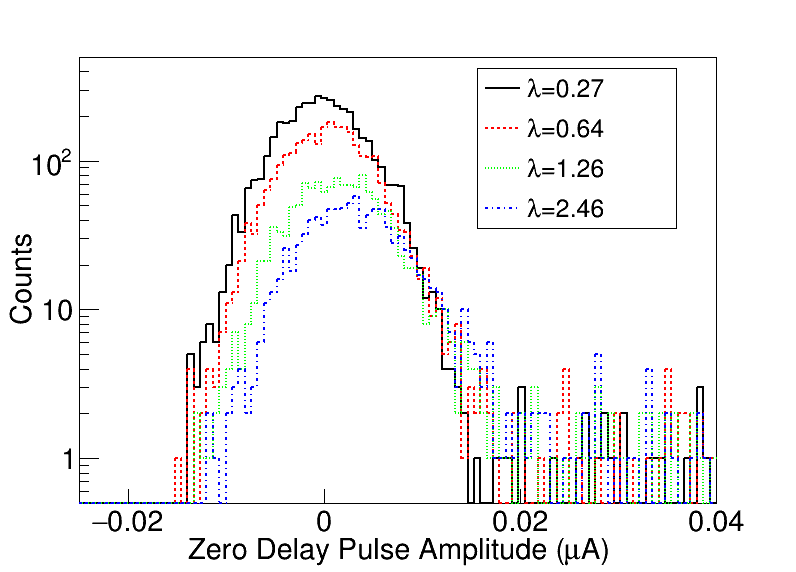}
    \caption{Distribution of $A_{\text{OF0}}$ values for events near the zeroth peak at different $\lambda$ values in the NF-C1 detector. Events near zero amplitude correspond to those without NTL amplification, while events to the right result from partial NTL amplification due to the charge trapping/impact ionization effect \cite{improved_CTII_model}.}
    \label{fig:zeroth_peak_offset}
\end{figure}

 The $\bar{E}_\text{offset}$ was quantified by analyzing multiple low-intensity LED datasets. At low flash intensities, some events may contain no photons reaching the detectors due to Poisson distribution; or the e$^-$h$^+$ pairs from incident photons may all undergo surface trapping. These events populated the zeroth peak with a mean value showing an offset with respect to zero energy, as shown in Fig.~\ref{fig:zeroth_peak_offset}. We measured the amplitude of pulses in the zeroth peak using $A_\text{OF0}$, and could model the offset of the zeroth peak at various LED intensities using
\begin{equation} 
\label{eq:peak_offset_vs_lambda}
    \bar{A}_{\rm offset} = m\lambda + l,
\end{equation}
where the linear relation in $m$ is motivated by the surface trapping hypothesis, and $l$ is associated with the cross-talk between the LED power lines and the detector readout. The pulse amplitudes were corrected to remove the effect of the cross-talk.

The energy equivalent of the surface trapping effect, $\bar{E}_{\text{offset}}$, was determined using
\begin{equation} 
    \bar{E}_{\text{offset}} 
    = \frac{152 \,\, \text{eV}}{(A^1_{\text{OF}} - A^0_{\text{OF0}})} \times m\lambda,
\label{eq:zeroth_offset}
\end{equation}
where 152~eV is the total phonon energy from an initial photon and the additional NTL gain, representing the energy difference between successive e$^-$h$^+$ peaks; $A^1_{\text{OF}}$ and $A^0_{\text{OF0}}$ are the averages of cross-talk corrected pulse amplitudes in the first and zeroth e$^-$h$^+$ peaks, and $m\lambda$ is the offset of the zeroth peak, measured in units of pulse amplitude, resulting from the surface trapping effect that is modeled Eqn.~\ref{eq:peak_offset_vs_lambda}. The prefactor in Eqn. \ref{eq:zeroth_offset} provides a preliminary calibration. The surface trapping effect in units of energy for a few LED datasets is shown in Table~\ref{tab:surface_trap_offset}.

\begin{table}[t]
\caption{Calibration function factors to convert $A_{\text{OF}}$ values to energy for the HV datasets.}
\begin{ruledtabular}
\begin{tabular}{lll}
\textrm{Detector} & $\alpha$ (keV/$\mu$A)  & $\beta$ (keV/$(\mu \text{A})^2$) \\
\colrule
\addlinespace[0.5ex] 
NF-H   & $1.90 \pm 0.04$ & $0.20 \pm 0.04 $  \\
NF-C1  & $0.74 \pm 0.01$ & $0.014  \pm 0.006 $  \\
NF-C2  & $1.11 \pm 0.02$ & $0.027 \pm 0.008$  \\
\end{tabular}
\end{ruledtabular}
\label{HV_calib_info}
\end{table}

\begin{figure}
\centering
    \includegraphics[width=\linewidth]{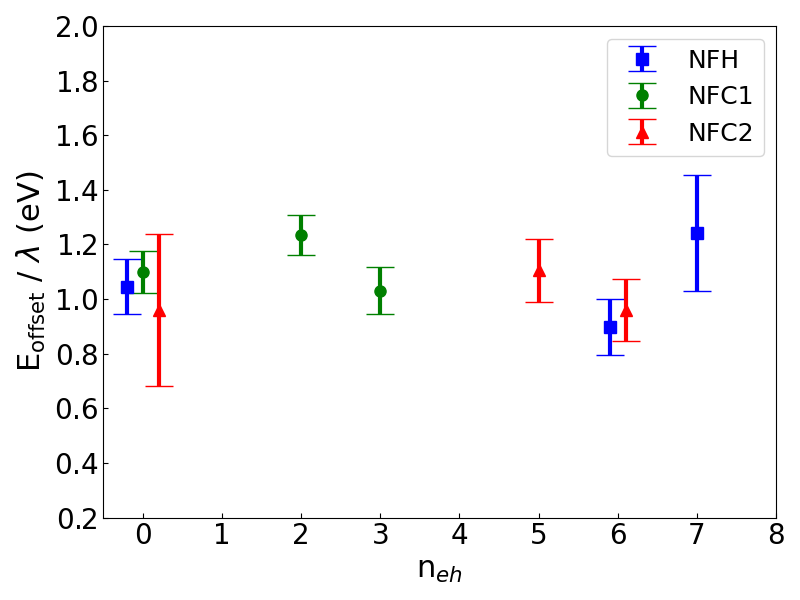}
  \caption{The ratio of the peak offset to $\lambda$ for various e$^-$h$^+$ pair peaks ($n_{\mathrm{eh}}$) in the NF-H, NF-C1, and NF-C2 detectors. The LED calibration data were taken separately for each detector, and the LED flash intensity was varied across detectors, resulting in different groups of available e$^-$h$^+$ peaks. A small x-axis offset is applied to the zeroth and sixth e$^-$h$^+$ pair peaks for the NF-H and NF-C2 detectors to improve readability.}
  \label{surface_trapping_effect}
\end{figure}

We validated the assumption of a constant energy offset in the LED calibration function
using experimental data. By increasing the intensity of LED flashes, we
confirmed that higher-order e$^{-}$h$^{+}$ peaks exhibit energy shifts consistent with that of the zeroth
peak. This comparison of energy shifts was carried out across different detectors, LED
sub-datasets, and e$^{-}$h$^{+}$ pair peaks, as shown in Fig.~\ref{surface_trapping_effect}. The consistency of the energy shifts between detectors is aligned with the surface trapping hypothesis~\cite{improved_CTII_model}, given that the LEDs illuminated the aluminum grid side of the detectors through pinholes, which have nearly identical characteristics for all detectors.

The consistency of the detector responses between the LED and Compton calibration datasets was
evaluated by comparing the HV calibrations in each setup. The biasing currents required to maintain the TES at 40\% of the normal resistance was different between the setups, resulting in changes in the responses of the detectors. We determined a constant scaling factor to equate the pulse amplitude of $\sim$150 eV events, before and after the installation of LED modules, while operating in HV mode. To quantify the uncertainty of the scaling factor for each detector, multiple LED datasets with TES biases surrounding the nominal value were collected. Linear corrections were determined to align the pulse amplitudes of their e$^-$h$^+$ peaks. After matching each e$^-$h$^+$ peak, the deviations in the energy of the other peaks were recorded. The largest deviation, on the order of 10 eV in each detector, was identified as the primary source of systematic uncertainty in the LED calibration.

\begin{figure*}[t]
    \centering
    \includegraphics[width=0.48\linewidth]{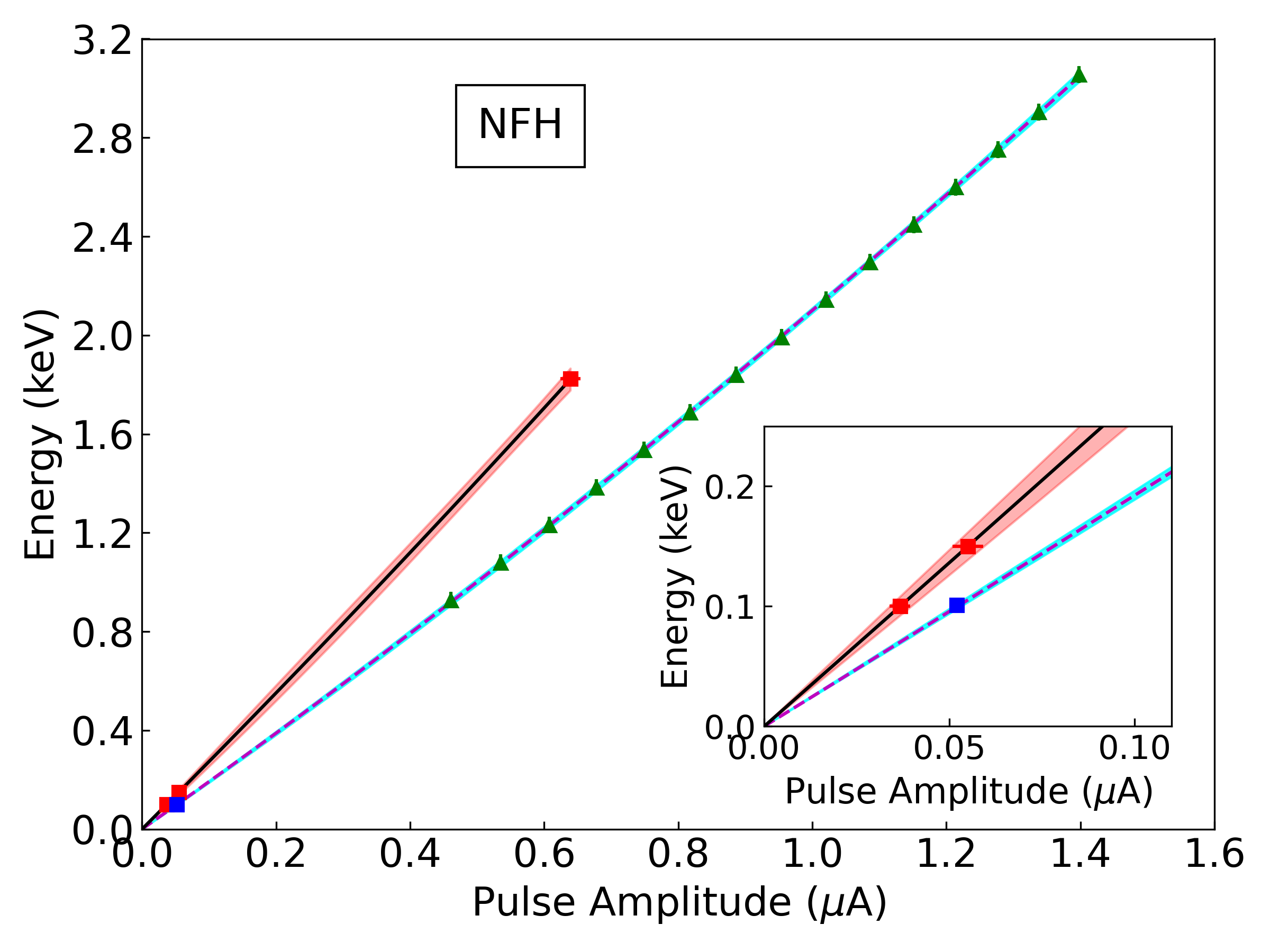}
    \includegraphics[width=0.48\linewidth]{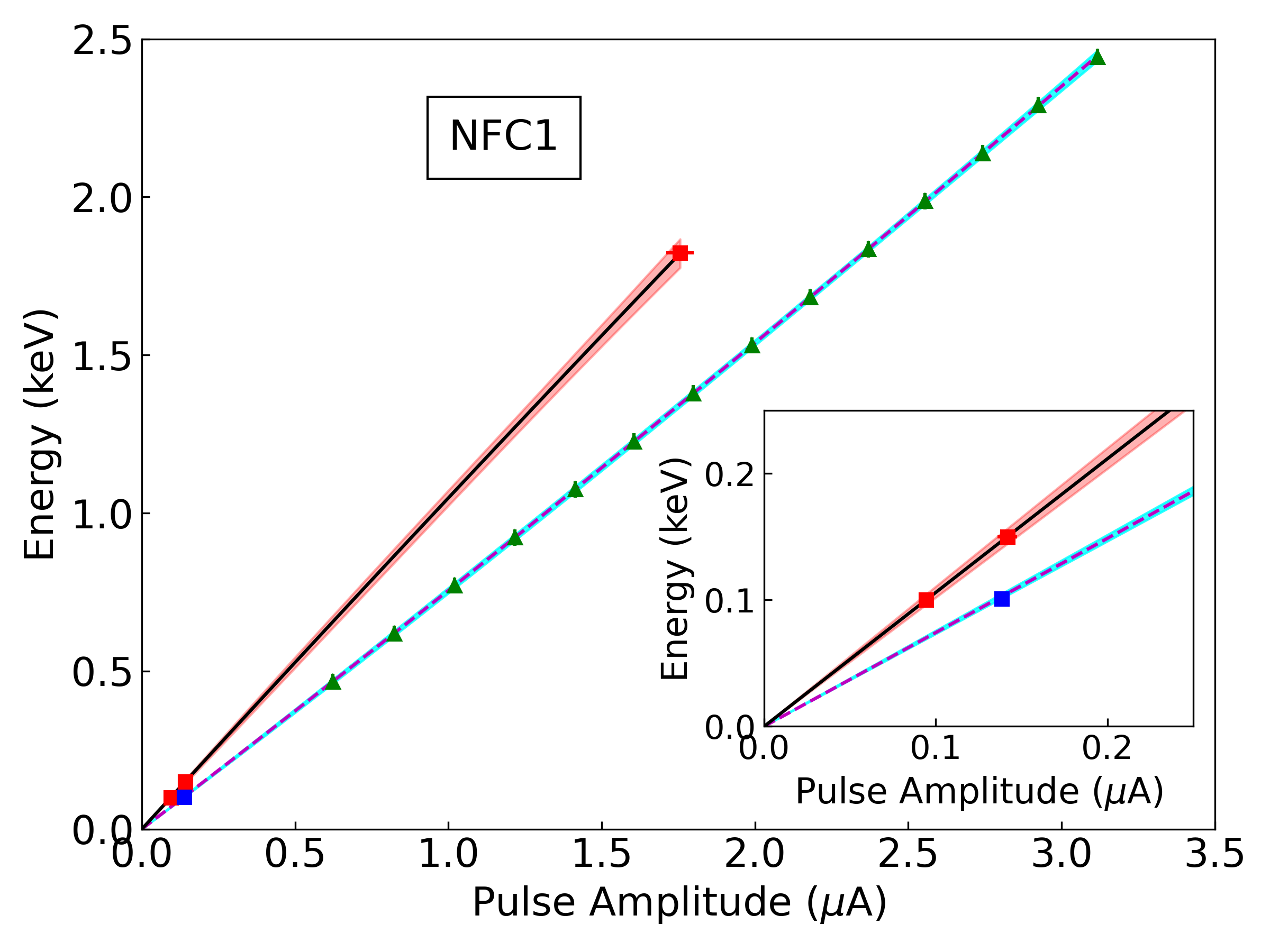}\\
    \hspace{80pt}
    \includegraphics[width=0.73\linewidth]{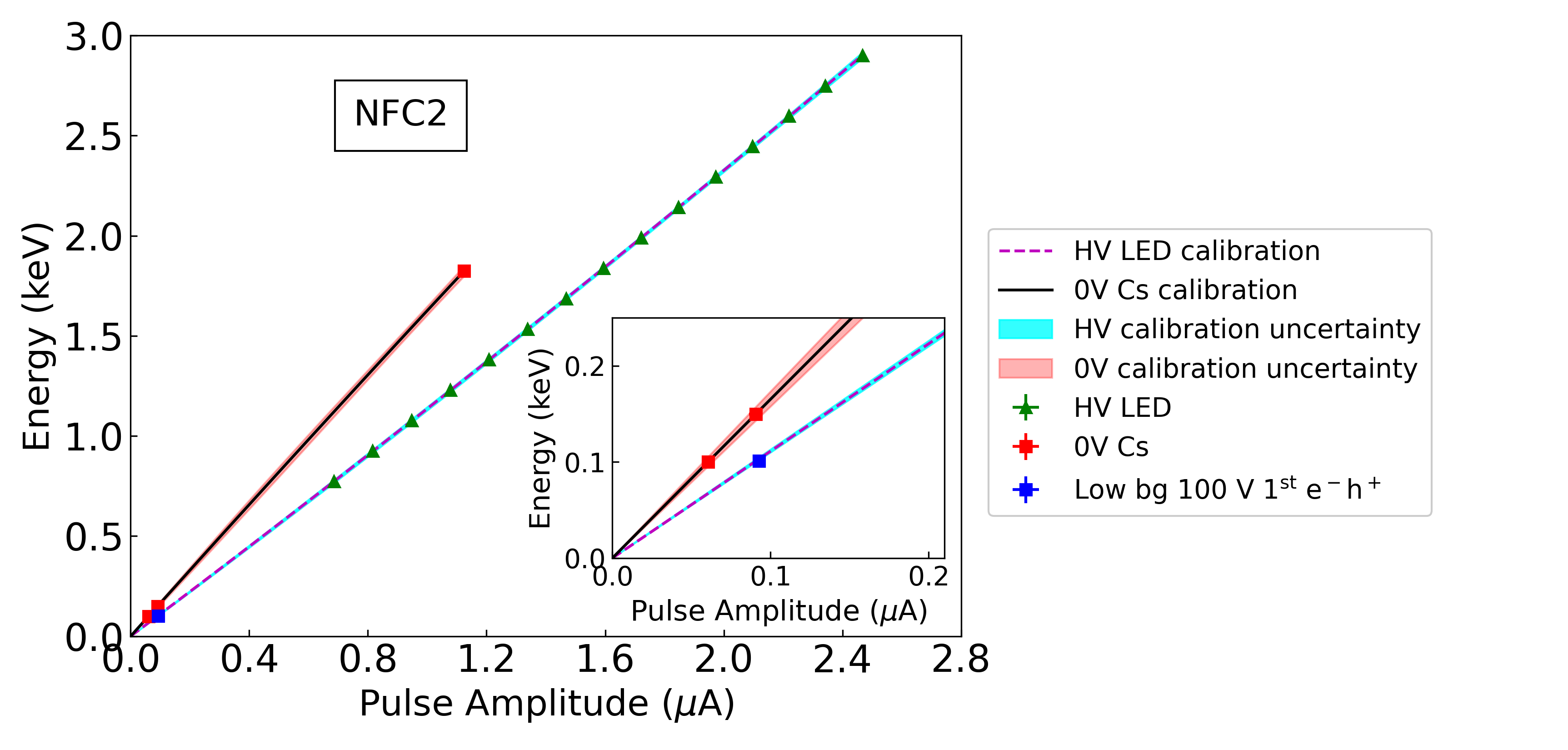}
  \caption{\label{final_calibration} Energy calibration curves at 0~V and 150~V (HV) bias conditions for NF-H, NF-C1, and NF-C2. A difference in detector response between the 0~V and HV calibration is evident. The inset plots show the low-energy region of the calibration curves.
 }
\end{figure*}

The average pulse amplitude of the e$^-$h$^+$ pair peaks, after applying cross talk and detector response corrections, follows a quadratic calibration function that is given in Eqn.~\ref{eqn:non-linear}. The coefficients of the HV LED calibration functions are provided in Table~\ref{HV_calib_info}. The HV LED calibration and the 0~V Compton calibration functions are shown in Fig.~\ref{final_calibration}. 

\section{Comparison of Calibration at 0~V and High Voltage Bias}
\label{sec:compare}

We calibrated the detectors at two bias voltages: (a) using Compton steps at 0~V, and (b) using e$^-$h$^+$-pair peaks induced by optical photons at 150~V. The pulse amplitude is calibrated to the interaction energy at 0~V, with the additional NTL gain considered in the HV operation. The calibration factor at HV is approximately 30\% smaller than the 0 V calibration factor. This difference implies that for the same deposited energy, the detector response in the 0 V calibration is approximately 30\% weaker than expected based on the HV calibration, as shown in Fig.~\ref{final_calibration}. The figures include an additional calibration point at approximately 100~eV~\cite {HVeVR3}, obtained from the first $e^-$h$^+$ pair peak in the background data collected while operating at a 100~V bias.

The difference in the calibration was measured by scaling the HV calibration function with a free multiplicative factor to fit the $0~\mathrm{V}$ calibration points. To determine the systematic uncertainty from the LED calibration, LED calibration curves were sampled within their uncertainty while accounting for correlations between calibration parameters. The sampled curves were fitted to the 0~V calibration points using the scaling parameter. The systematic uncertainty was determined from the spread in the values of the scaling parameter and was combined in quadrature with the statistical uncertainty from fitting the primary LED calibration curve. The percentage difference between the 0~V and HV calibrations, relative to the 0~V calibration, is summarized in Table~\ref{tab:0V_HV_discrepancy}.

A similar calibration difference between the 0~V and HV operations of detector NFC1 has been observed previously, and several hypotheses have been proposed to explain it~\cite{HVeV_characterization}. At 0~V, this study used energy signatures from bulk electron interactions, in contrast with keV-scale surface interactions produced by $^{55}$Fe X-rays in the previous study. The similarity in calibration differences between the two studies may offer insights about the physics. A more detailed investigation and discussion of the underlying mechanism are deferred to future work.

\section{conclusion and outlook}
We calibrated the SuperCDMS Si HVeV detectors using L-shells (0.1~keV and 0.15~keV) and K-shell (1.8~keV) Compton steps at 0~V detector bias. This study presents the first measurement of L-shell Compton steps in cryogenic silicon calorimeters. We also performed an energy calibration up to a few keV using bursts of 2~eV optical photons from LEDs at 150~V detector bias. Our results show an approximately 30\% weaker detector response at 0 V crystal bias compared to high-voltage operation for the same expected phonon energy. A future measurement with additional Compton calibration data may help to validate the treatment of the core holes in FEFF calculations. The development of the calibration scheme with Compton steps for cryogenic calorimeters will be crucial for experiments such as SuperCDMS SNOLAB to take full advantage of the low energy threshold and excellent energy resolution of the detectors in the search for low-mass dark matter.

\begin{table}
\caption{Comparison of the percentage differences between the 0~V and HV calibration relative to the 0~V calibration, obtained from a one-parameter fit for different HVeV detectors.}
\begin{ruledtabular}
\begin{tabular}{lccc}
\textrm{Detector} & \textrm{Calibration difference (\%)} & \multicolumn{2}{c}{\textrm{Uncertainty (\%)}} \\
 & & \textrm{Stat} & \textrm{Sys} \\
\colrule
\addlinespace[0.5ex] 
NF-H  &  29 $\pm$ 2 & 1.7 &  0.7 \\
NF-C1 &  27 $\pm$ 2 & 1.6 &  0.7 \\
NF-C2 &  30 $\pm$  1 & 1.0 &  0.5 \\
\end{tabular}
\end{ruledtabular}
\label{tab:0V_HV_discrepancy}
\end{table}

\section{Acknowledgments}
We are grateful to Aman Singal and Rouven Essig for their helpful suggestions on the numerical calculations of the Compton scattering differential cross section. We thank Alvaro Chavarria for sharing insights on the FEFF calculations performed by the DAMIC-M collaboration, and John J. Rehr for his guidance with the FEFF calculations. This research was enabled in part by support provided by Compute Ontario (computeontario.ca) and the Digital Research Alliance of Canada (alliancecan.ca). Funding and support were received from the National Science Foundation, the U.S. Department of Energy (DOE), Fermilab URA Visiting Scholar Grant No. 15-S-33, NSERC Canada, the Canada First Excellence Research  Fund, the Arthur B. McDonald Institute (Canada),  the Department of Atomic Energy Government of India (DAE), J. C. Bose Fellowship grant of the Anusandhan National Research Foundation (ANRF, India) and the DFG (Germany) - Project No. 420484612 and under Germany’s Excellence Strategy - EXC 2121 ``Quantum Universe" – 390833306 and the Marie-Curie program - Contract No. 101104484. Fermilab is operated by Fermi Forward Discovery Group, LLC,  SLAC is operated by Stanford University, and PNNL is operated by the Battelle Memorial Institute for the U.S. Department of Energy under contracts DE-AC02-37407CH11359, DE-AC02-76SF00515, and DE-AC05-76RL01830, respectively.

\bibliography{main}

\begin{thebibliography}{45}%
\makeatletter
\providecommand \@ifxundefined [1]{%
 \@ifx{#1\undefined}
}%
\providecommand \@ifnum [1]{%
 \ifnum #1\expandafter \@firstoftwo
 \else \expandafter \@secondoftwo
 \fi
}%
\providecommand \@ifx [1]{%
 \ifx #1\expandafter \@firstoftwo
 \else \expandafter \@secondoftwo
 \fi
}%
\providecommand \natexlab [1]{#1}%
\providecommand \enquote  [1]{``#1''}%
\providecommand \bibnamefont  [1]{#1}%
\providecommand \bibfnamefont [1]{#1}%
\providecommand \citenamefont [1]{#1}%
\providecommand \href@noop [0]{\@secondoftwo}%
\providecommand \href [0]{\begingroup \@sanitize@url \@href}%
\providecommand \@href[1]{\@@startlink{#1}\@@href}%
\providecommand \@@href[1]{\endgroup#1\@@endlink}%
\providecommand \@sanitize@url [0]{\catcode `\\12\catcode `\$12\catcode `\&12\catcode `\#12\catcode `\^12\catcode `\_12\catcode `\%12\relax}%
\providecommand \@@startlink[1]{}%
\providecommand \@@endlink[0]{}%
\providecommand \url  [0]{\begingroup\@sanitize@url \@url }%
\providecommand \@url [1]{\endgroup\@href {#1}{\urlprefix }}%
\providecommand \urlprefix  [0]{URL }%
\providecommand \Eprint [0]{\href }%
\providecommand \doibase [0]{http://dx.doi.org/}%
\providecommand \selectlanguage [0]{\@gobble}%
\providecommand \bibinfo  [0]{\@secondoftwo}%
\providecommand \bibfield  [0]{\@secondoftwo}%
\providecommand \translation [1]{[#1]}%
\providecommand \BibitemOpen [0]{}%
\providecommand \bibitemStop [0]{}%
\providecommand \bibitemNoStop [0]{.\EOS\space}%
\providecommand \EOS [0]{\spacefactor3000\relax}%
\providecommand \BibitemShut  [1]{\csname bibitem#1\endcsname}%
\let\auto@bib@innerbib\@empty
\bibitem [{\citenamefont {Albakry}\ \emph {et~al.}()\citenamefont {Albakry} \emph {et~al.}}]{albakry2022strategy}%
  \BibitemOpen
  \bibfield  {author} {\bibinfo {author} {\bibfnamefont {M.~F.}\ \bibnamefont {Albakry}} \emph {et~al.},\ }\Eprint {http://arxiv.org/abs/2203.08463} {arXiv:2203.08463} \BibitemShut {NoStop}%
\bibitem [{\citenamefont {Essig}\ \emph {et~al.}({\natexlab{a}})\citenamefont {Essig} \emph {et~al.}}]{essig2022snowmass2021_1}%
  \BibitemOpen
  \bibfield  {author} {\bibinfo {author} {\bibfnamefont {R.}~\bibnamefont {Essig}} \emph {et~al.},\ }\Eprint {http://arxiv.org/abs/2203.08297} {arXiv:2203.08297} \BibitemShut {NoStop}%
\bibitem [{\citenamefont {Essig}\ \emph {et~al.}({\natexlab{b}})\citenamefont {Essig}, \citenamefont {Kahn}, \citenamefont {Knapen}, \citenamefont {Ringwald},\ and\ \citenamefont {Toro}}]{essig2022snowmass2021_2}%
  \BibitemOpen
  \bibfield  {author} {\bibinfo {author} {\bibfnamefont {R.}~\bibnamefont {Essig}}, \bibinfo {author} {\bibfnamefont {Y.}~\bibnamefont {Kahn}}, \bibinfo {author} {\bibfnamefont {S.}~\bibnamefont {Knapen}}, \bibinfo {author} {\bibfnamefont {A.}~\bibnamefont {Ringwald}}, \ and\ \bibinfo {author} {\bibfnamefont {N.}~\bibnamefont {Toro}},\ }\Eprint {http://arxiv.org/abs/2203.10089} {arXiv:2203.10089} \BibitemShut {NoStop}%
\bibitem [{\citenamefont {Anthony-Petersen}\ \emph {et~al.}(2024)\citenamefont {Anthony-Petersen} \emph {et~al.}}]{anthony2023applying}%
  \BibitemOpen
  \bibfield  {author} {\bibinfo {author} {\bibfnamefont {R.}~\bibnamefont {Anthony-Petersen}} \emph {et~al.} (\bibinfo {collaboration} {SPICE/HeRALD Collaboration}),\ }\href {\doibase 10.1103/PhysRevD.110.072006} {\bibfield  {journal} {\bibinfo  {journal} {Physical Review D}\ }\textbf {\bibinfo {volume} {110}},\ \bibinfo {pages} {072006} (\bibinfo {year} {2024})}\BibitemShut {NoStop}%
\bibitem [{\citenamefont {Tiffenberg}\ \emph {et~al.}(2017)\citenamefont {Tiffenberg}, \citenamefont {Sofo-Haro}, \citenamefont {Drlica-Wagner}, \citenamefont {Essig}, \citenamefont {Guardincerri}, \citenamefont {Holland}, \citenamefont {Volansky},\ and\ \citenamefont {Yu}}]{PhysRevLett.119.131802}%
  \BibitemOpen
  \bibfield  {author} {\bibinfo {author} {\bibfnamefont {J.}~\bibnamefont {Tiffenberg}}, \bibinfo {author} {\bibfnamefont {M.}~\bibnamefont {Sofo-Haro}}, \bibinfo {author} {\bibfnamefont {A.}~\bibnamefont {Drlica-Wagner}}, \bibinfo {author} {\bibfnamefont {R.}~\bibnamefont {Essig}}, \bibinfo {author} {\bibfnamefont {Y.}~\bibnamefont {Guardincerri}}, \bibinfo {author} {\bibfnamefont {S.}~\bibnamefont {Holland}}, \bibinfo {author} {\bibfnamefont {T.}~\bibnamefont {Volansky}}, \ and\ \bibinfo {author} {\bibfnamefont {T.-T.}\ \bibnamefont {Yu}},\ }\href {\doibase 10.1103/PhysRevLett.119.131802} {\bibfield  {journal} {\bibinfo  {journal} {Physical Review Letters}\ }\textbf {\bibinfo {volume} {119}},\ \bibinfo {pages} {131802} (\bibinfo {year} {2017})}\BibitemShut {NoStop}%
\bibitem [{\citenamefont {Hong}\ \emph {et~al.}(2020)\citenamefont {Hong}, \citenamefont {Ren}, \citenamefont {Kurinsky}, \citenamefont {Figueroa-Feliciano}, \citenamefont {Wills}, \citenamefont {Ganjam}, \citenamefont {Mahapatra}, \citenamefont {Mirabolfathi}, \citenamefont {Nebolsky}, \citenamefont {Pinckney},\ and\ \citenamefont {Platt}}]{HONG2020163757}%
  \BibitemOpen
  \bibfield  {author} {\bibinfo {author} {\bibfnamefont {Z.}~\bibnamefont {Hong}}, \bibinfo {author} {\bibfnamefont {R.}~\bibnamefont {Ren}}, \bibinfo {author} {\bibfnamefont {N.}~\bibnamefont {Kurinsky}}, \bibinfo {author} {\bibfnamefont {E.}~\bibnamefont {Figueroa-Feliciano}}, \bibinfo {author} {\bibfnamefont {L.}~\bibnamefont {Wills}}, \bibinfo {author} {\bibfnamefont {S.}~\bibnamefont {Ganjam}}, \bibinfo {author} {\bibfnamefont {R.}~\bibnamefont {Mahapatra}}, \bibinfo {author} {\bibfnamefont {N.}~\bibnamefont {Mirabolfathi}}, \bibinfo {author} {\bibfnamefont {B.}~\bibnamefont {Nebolsky}}, \bibinfo {author} {\bibfnamefont {H.~D.}\ \bibnamefont {Pinckney}}, \ and\ \bibinfo {author} {\bibfnamefont {M.}~\bibnamefont {Platt}},\ }\href {\doibase https://doi.org/10.1016/j.nima.2020.163757} {\bibfield  {journal} {\bibinfo  {journal} {Nuclear Instruments and Methods in Physics Research Section A: Accelerators, Spectrometers, Detectors and Associated Equipment}\ }\textbf {\bibinfo {volume} {963}},\ \bibinfo {pages}
  {163757} (\bibinfo {year} {2020})}\BibitemShut {NoStop}%
\bibitem [{\citenamefont {Ren}\ \emph {et~al.}(2021)\citenamefont {Ren}, \citenamefont {Bathurst}, \citenamefont {Chang}, \citenamefont {Chen}, \citenamefont {Fink}, \citenamefont {Hong}, \citenamefont {Kurinsky}, \citenamefont {Mast}, \citenamefont {Mishra}, \citenamefont {Novati} \emph {et~al.}}]{HVeV_characterization}%
  \BibitemOpen
  \bibfield  {author} {\bibinfo {author} {\bibfnamefont {R.}~\bibnamefont {Ren}}, \bibinfo {author} {\bibfnamefont {C.}~\bibnamefont {Bathurst}}, \bibinfo {author} {\bibfnamefont {Y.}~\bibnamefont {Chang}}, \bibinfo {author} {\bibfnamefont {R.}~\bibnamefont {Chen}}, \bibinfo {author} {\bibfnamefont {C.}~\bibnamefont {Fink}}, \bibinfo {author} {\bibfnamefont {Z.}~\bibnamefont {Hong}}, \bibinfo {author} {\bibfnamefont {N.}~\bibnamefont {Kurinsky}}, \bibinfo {author} {\bibfnamefont {N.}~\bibnamefont {Mast}}, \bibinfo {author} {\bibfnamefont {N.}~\bibnamefont {Mishra}}, \bibinfo {author} {\bibfnamefont {V.}~\bibnamefont {Novati}},  \emph {et~al.},\ }\href {\doibase 10.1103/PhysRevD.104.032010} {\bibfield  {journal} {\bibinfo  {journal} {Physical Review D}\ }\textbf {\bibinfo {volume} {104}},\ \bibinfo {pages} {032010} (\bibinfo {year} {2021})}\BibitemShut {NoStop}%
\bibitem [{\citenamefont {Agnese}\ \emph {et~al.}(2018)\citenamefont {Agnese} \emph {et~al.}}]{HVeVR1}%
  \BibitemOpen
  \bibfield  {author} {\bibinfo {author} {\bibfnamefont {R.}~\bibnamefont {Agnese}} \emph {et~al.} (\bibinfo {collaboration} {SuperCDMS Collaboration}),\ }\href {\doibase 10.1103/PhysRevLett.121.051301} {\bibfield  {journal} {\bibinfo  {journal} {Physical Review Letters}\ }\textbf {\bibinfo {volume} {121}},\ \bibinfo {pages} {051301} (\bibinfo {year} {2018})}\BibitemShut {NoStop}%
\bibitem [{\citenamefont {Amaral}\ \emph {et~al.}(2020)\citenamefont {Amaral} \emph {et~al.}}]{HVeVR2}%
  \BibitemOpen
  \bibfield  {author} {\bibinfo {author} {\bibfnamefont {D.}~\bibnamefont {Amaral}} \emph {et~al.} (\bibinfo {collaboration} {SuperCDMS Collaboration}),\ }\href {\doibase 10.1103/PhysRevD.102.091101} {\bibfield  {journal} {\bibinfo  {journal} {Physical Review D}\ }\textbf {\bibinfo {volume} {102}},\ \bibinfo {pages} {091101} (\bibinfo {year} {2020})}\BibitemShut {NoStop}%
\bibitem [{\citenamefont {Albakry}\ \emph {et~al.}(2022)\citenamefont {Albakry} \emph {et~al.}}]{Low_energy_events_HVeV}%
  \BibitemOpen
  \bibfield  {author} {\bibinfo {author} {\bibfnamefont {M.}~\bibnamefont {Albakry}} \emph {et~al.} (\bibinfo {collaboration} {SuperCDMS Collaboration}),\ }\href {\doibase 10.1103/PhysRevD.105.112006} {\bibfield  {journal} {\bibinfo  {journal} {Physical Review D}\ }\textbf {\bibinfo {volume} {105}},\ \bibinfo {pages} {112006} (\bibinfo {year} {2022})}\BibitemShut {NoStop}%
\bibitem [{\citenamefont {Albakry}\ \emph {et~al.}(2025)\citenamefont {Albakry} \emph {et~al.}}]{HVeVR3}%
  \BibitemOpen
  \bibfield  {author} {\bibinfo {author} {\bibfnamefont {M.~F.}\ \bibnamefont {Albakry}} \emph {et~al.} (\bibinfo {collaboration} {SuperCDMS Collaboration}),\ }\href {\doibase 10.1103/PhysRevD.111.012006} {\bibfield  {journal} {\bibinfo  {journal} {Physical Review D}\ }\textbf {\bibinfo {volume} {111}},\ \bibinfo {pages} {012006} (\bibinfo {year} {2025})}\BibitemShut {NoStop}%
\bibitem [{\citenamefont {Norcini}\ \emph {et~al.}(2022)\citenamefont {Norcini} \emph {et~al.}}]{norcini2022precision}%
  \BibitemOpen
  \bibfield  {author} {\bibinfo {author} {\bibfnamefont {D.}~\bibnamefont {Norcini}} \emph {et~al.} (\bibinfo {collaboration} {DAMIC-M Collaboration}),\ }\href {\doibase 10.1103/PhysRevD.106.092001} {\bibfield  {journal} {\bibinfo  {journal} {Physical Review D}\ }\textbf {\bibinfo {volume} {106}},\ \bibinfo {pages} {092001} (\bibinfo {year} {2022})}\BibitemShut {NoStop}%
\bibitem [{\citenamefont {Adamson}\ \emph {et~al.}(2015)\citenamefont {Adamson}, \citenamefont {Anghel}, \citenamefont {Aurisano}, \citenamefont {Barr}, \citenamefont {Bishai}, \citenamefont {Blake}, \citenamefont {Bock}, \citenamefont {Bogert}, \citenamefont {Cao}, \citenamefont {Castromonte} \emph {et~al.}}]{adamson2015observation}%
  \BibitemOpen
  \bibfield  {author} {\bibinfo {author} {\bibfnamefont {P.}~\bibnamefont {Adamson}}, \bibinfo {author} {\bibfnamefont {I.}~\bibnamefont {Anghel}}, \bibinfo {author} {\bibfnamefont {A.}~\bibnamefont {Aurisano}}, \bibinfo {author} {\bibfnamefont {G.}~\bibnamefont {Barr}}, \bibinfo {author} {\bibfnamefont {M.}~\bibnamefont {Bishai}}, \bibinfo {author} {\bibfnamefont {A.}~\bibnamefont {Blake}}, \bibinfo {author} {\bibfnamefont {G.}~\bibnamefont {Bock}}, \bibinfo {author} {\bibfnamefont {D.}~\bibnamefont {Bogert}}, \bibinfo {author} {\bibfnamefont {S.}~\bibnamefont {Cao}}, \bibinfo {author} {\bibfnamefont {C.}~\bibnamefont {Castromonte}},  \emph {et~al.},\ }\href {\doibase 10.1103/PhysRevD.91.112006} {\bibfield  {journal} {\bibinfo  {journal} {Physical Review D}\ }\textbf {\bibinfo {volume} {91}},\ \bibinfo {pages} {112006} (\bibinfo {year} {2015})}\BibitemShut {NoStop}%
\bibitem [{\citenamefont {Wilson}\ \emph {et~al.}(2024)\citenamefont {Wilson}, \citenamefont {Zaytsev}, \citenamefont {von Krosigk}, \citenamefont {Alkhatib}, \citenamefont {Buchanan}, \citenamefont {Chen}, \citenamefont {Diamond}, \citenamefont {Figueroa-Feliciano}, \citenamefont {Harms}, \citenamefont {Hong} \emph {et~al.}}]{improved_CTII_model}%
  \BibitemOpen
  \bibfield  {author} {\bibinfo {author} {\bibfnamefont {M.}~\bibnamefont {Wilson}}, \bibinfo {author} {\bibfnamefont {A.}~\bibnamefont {Zaytsev}}, \bibinfo {author} {\bibfnamefont {B.}~\bibnamefont {von Krosigk}}, \bibinfo {author} {\bibfnamefont {I.}~\bibnamefont {Alkhatib}}, \bibinfo {author} {\bibfnamefont {M.}~\bibnamefont {Buchanan}}, \bibinfo {author} {\bibfnamefont {R.}~\bibnamefont {Chen}}, \bibinfo {author} {\bibfnamefont {M.}~\bibnamefont {Diamond}}, \bibinfo {author} {\bibfnamefont {E.}~\bibnamefont {Figueroa-Feliciano}}, \bibinfo {author} {\bibfnamefont {S.}~\bibnamefont {Harms}}, \bibinfo {author} {\bibfnamefont {Z.}~\bibnamefont {Hong}},  \emph {et~al.},\ }\href {\doibase 10.1103/PhysRevD.109.112018} {\bibfield  {journal} {\bibinfo  {journal} {Physical Review D}\ }\textbf {\bibinfo {volume} {109}},\ \bibinfo {pages} {112018} (\bibinfo {year} {2024})}\BibitemShut {NoStop}%
\bibitem [{\citenamefont {Agostinelli}\ \emph {et~al.}(2003)\citenamefont {Agostinelli} \emph {et~al.}}]{GEANT4:2002zbu}%
  \BibitemOpen
  \bibfield  {author} {\bibinfo {author} {\bibfnamefont {S.}~\bibnamefont {Agostinelli}} \emph {et~al.} (\bibinfo {collaboration} {GEANT4 Collaboration}),\ }\href {\doibase 10.1016/S0168-9002(03)01368-8} {\bibfield  {journal} {\bibinfo  {journal} {Nuclear Instruments and Methods in Physics Research Section A: Accelerators, Spectrometers, Detectors and Associated Equipment}\ }\textbf {\bibinfo {volume} {506}},\ \bibinfo {pages} {250} (\bibinfo {year} {2003})}\BibitemShut {NoStop}%
\bibitem [{\citenamefont {Brown}\ \emph {et~al.}(2014)\citenamefont {Brown}, \citenamefont {Dimmock}, \citenamefont {Gillam},\ and\ \citenamefont {Paganin}}]{f8bba843a6e44af48b77e2489f067559}%
  \BibitemOpen
  \bibfield  {author} {\bibinfo {author} {\bibfnamefont {J.}~\bibnamefont {Brown}}, \bibinfo {author} {\bibfnamefont {M.}~\bibnamefont {Dimmock}}, \bibinfo {author} {\bibfnamefont {J.}~\bibnamefont {Gillam}}, \ and\ \bibinfo {author} {\bibfnamefont {D.}~\bibnamefont {Paganin}},\ }\href {\doibase 10.1016/j.nimb.2014.07.042} {\bibfield  {journal} {\bibinfo  {journal} {Nuclear Instruments and Methods in Physics Research, Section B: Beam Interactions with Materials and Atoms}\ }\textbf {\bibinfo {volume} {338}},\ \bibinfo {pages} {77 } (\bibinfo {year} {2014})}\BibitemShut {NoStop}%
\bibitem [{\citenamefont {Du~Mond}(1929)}]{PhysRev.33.643}%
  \BibitemOpen
  \bibfield  {author} {\bibinfo {author} {\bibfnamefont {J.~W.~M.}\ \bibnamefont {Du~Mond}},\ }\href {\doibase 10.1103/PhysRev.33.643} {\bibfield  {journal} {\bibinfo  {journal} {Physical Review}\ }\textbf {\bibinfo {volume} {33}},\ \bibinfo {pages} {643} (\bibinfo {year} {1929})}\BibitemShut {NoStop}%
\bibitem [{\citenamefont {Eisenberger}\ and\ \citenamefont {Platzman}(1970)}]{PhysRevA.2.415}%
  \BibitemOpen
  \bibfield  {author} {\bibinfo {author} {\bibfnamefont {P.}~\bibnamefont {Eisenberger}}\ and\ \bibinfo {author} {\bibfnamefont {P.~M.}\ \bibnamefont {Platzman}},\ }\href {\doibase 10.1103/PhysRevA.2.415} {\bibfield  {journal} {\bibinfo  {journal} {Physical Review A}\ }\textbf {\bibinfo {volume} {2}},\ \bibinfo {pages} {415} (\bibinfo {year} {1970})}\BibitemShut {NoStop}%
\bibitem [{\citenamefont {Ribberfors}(1975)}]{PhysRevB.12.3136}%
  \BibitemOpen
  \bibfield  {author} {\bibinfo {author} {\bibfnamefont {R.}~\bibnamefont {Ribberfors}},\ }\href {\doibase 10.1103/PhysRevB.12.3136} {\bibfield  {journal} {\bibinfo  {journal} {Physical Review B}\ }\textbf {\bibinfo {volume} {12}},\ \bibinfo {pages} {3136} (\bibinfo {year} {1975})}\BibitemShut {NoStop}%
\bibitem [{\citenamefont {Kas}\ \emph {et~al.}(2021)\citenamefont {Kas}, \citenamefont {Vila}, \citenamefont {Pemmaraju}, \citenamefont {Tan},\ and\ \citenamefont {Rehr}}]{kas2021advanced}%
  \BibitemOpen
  \bibfield  {author} {\bibinfo {author} {\bibfnamefont {J.}~\bibnamefont {Kas}}, \bibinfo {author} {\bibfnamefont {F.}~\bibnamefont {Vila}}, \bibinfo {author} {\bibfnamefont {C.}~\bibnamefont {Pemmaraju}}, \bibinfo {author} {\bibfnamefont {T.}~\bibnamefont {Tan}}, \ and\ \bibinfo {author} {\bibfnamefont {J.}~\bibnamefont {Rehr}},\ }\href {\doibase https://doi.org/10.1107/S1600577521008614} {\bibfield  {journal} {\bibinfo  {journal} {Journal of Synchrotron Radiation}\ }\textbf {\bibinfo {volume} {28}},\ \bibinfo {pages} {1801} (\bibinfo {year} {2021})}\BibitemShut {NoStop}%
\bibitem [{\citenamefont {Rehr}\ \emph {et~al.}(2009)\citenamefont {Rehr}, \citenamefont {Kas}, \citenamefont {Prange}, \citenamefont {Sorini}, \citenamefont {Takimoto},\ and\ \citenamefont {Vila}}]{REHR2009548}%
  \BibitemOpen
  \bibfield  {author} {\bibinfo {author} {\bibfnamefont {J.~J.}\ \bibnamefont {Rehr}}, \bibinfo {author} {\bibfnamefont {J.~J.}\ \bibnamefont {Kas}}, \bibinfo {author} {\bibfnamefont {M.~P.}\ \bibnamefont {Prange}}, \bibinfo {author} {\bibfnamefont {A.~P.}\ \bibnamefont {Sorini}}, \bibinfo {author} {\bibfnamefont {Y.}~\bibnamefont {Takimoto}}, \ and\ \bibinfo {author} {\bibfnamefont {F.}~\bibnamefont {Vila}},\ }\href {\doibase https://doi.org/10.1016/j.crhy.2008.08.004} {\bibfield  {journal} {\bibinfo  {journal} {Comptes Rendus Physique}\ }\textbf {\bibinfo {volume} {10}},\ \bibinfo {pages} {548} (\bibinfo {year} {2009})}\BibitemShut {NoStop}%
\bibitem [{\citenamefont {Rehr}\ and\ \citenamefont {Albers}(2000)}]{RevModPhys.72.621}%
  \BibitemOpen
  \bibfield  {author} {\bibinfo {author} {\bibfnamefont {J.~J.}\ \bibnamefont {Rehr}}\ and\ \bibinfo {author} {\bibfnamefont {R.~C.}\ \bibnamefont {Albers}},\ }\href {\doibase 10.1103/RevModPhys.72.621} {\bibfield  {journal} {\bibinfo  {journal} {Reviews of Modern Physics}\ }\textbf {\bibinfo {volume} {72}},\ \bibinfo {pages} {621} (\bibinfo {year} {2000})}\BibitemShut {NoStop}%
\bibitem [{\citenamefont {Rehr}\ \emph {et~al.}(2010)\citenamefont {Rehr}, \citenamefont {Kas}, \citenamefont {Vila}, \citenamefont {Prange},\ and\ \citenamefont {Jorissen}}]{feff4}%
  \BibitemOpen
  \bibfield  {author} {\bibinfo {author} {\bibfnamefont {J.~J.}\ \bibnamefont {Rehr}}, \bibinfo {author} {\bibfnamefont {J.~J.}\ \bibnamefont {Kas}}, \bibinfo {author} {\bibfnamefont {F.~D.}\ \bibnamefont {Vila}}, \bibinfo {author} {\bibfnamefont {M.~P.}\ \bibnamefont {Prange}}, \ and\ \bibinfo {author} {\bibfnamefont {K.}~\bibnamefont {Jorissen}},\ }\href {\doibase https://doi.org/10.1039/B926434E} {\bibfield  {journal} {\bibinfo  {journal} {Physical Chemistry Chemical Physics}\ }\textbf {\bibinfo {volume} {12}},\ \bibinfo {pages} {5503} (\bibinfo {year} {2010})}\BibitemShut {NoStop}%
\bibitem [{\citenamefont {Soininen}\ \emph {et~al.}(2005)\citenamefont {Soininen}, \citenamefont {Ankudinov},\ and\ \citenamefont {Rehr}}]{feff5}%
  \BibitemOpen
  \bibfield  {author} {\bibinfo {author} {\bibfnamefont {J.~A.}\ \bibnamefont {Soininen}}, \bibinfo {author} {\bibfnamefont {A.~L.}\ \bibnamefont {Ankudinov}}, \ and\ \bibinfo {author} {\bibfnamefont {J.~J.}\ \bibnamefont {Rehr}},\ }\href {\doibase 10.1103/PhysRevB.72.045136} {\bibfield  {journal} {\bibinfo  {journal} {Physical Review B}\ }\textbf {\bibinfo {volume} {72}},\ \bibinfo {pages} {045136} (\bibinfo {year} {2005})}\BibitemShut {NoStop}%
\bibitem [{\citenamefont {Massey}(1951)}]{16e7f618-c06b-3d10-8705-1086b218d827}%
  \BibitemOpen
  \bibfield  {author} {\bibinfo {author} {\bibfnamefont {F.~J.}\ \bibnamefont {Massey}},\ }\href {http://www.jstor.org/stable/2280095} {\bibfield  {journal} {\bibinfo  {journal} {Journal of the American Statistical Association}\ }\textbf {\bibinfo {volume} {46}},\ \bibinfo {pages} {68} (\bibinfo {year} {1951})}\BibitemShut {NoStop}%
\bibitem [{\citenamefont {Ravel}(2001)}]{Ravel:hf5152}%
  \BibitemOpen
  \bibfield  {author} {\bibinfo {author} {\bibfnamefont {B.}~\bibnamefont {Ravel}},\ }\href {\doibase https://doi.org/10.1107/S090904950001493X} {\bibfield  {journal} {\bibinfo  {journal} {Journal of synchrotron radiation}\ }\textbf {\bibinfo {volume} {8}},\ \bibinfo {pages} {314} (\bibinfo {year} {2001})}\BibitemShut {NoStop}%
\bibitem [{\citenamefont {Mattern}\ \emph {et~al.}(2012)\citenamefont {Mattern}, \citenamefont {Seidler}, \citenamefont {Kas}, \citenamefont {Pacold},\ and\ \citenamefont {Rehr}}]{PhysRevB.85.115135}%
  \BibitemOpen
  \bibfield  {author} {\bibinfo {author} {\bibfnamefont {B.~A.}\ \bibnamefont {Mattern}}, \bibinfo {author} {\bibfnamefont {G.~T.}\ \bibnamefont {Seidler}}, \bibinfo {author} {\bibfnamefont {J.~J.}\ \bibnamefont {Kas}}, \bibinfo {author} {\bibfnamefont {J.~I.}\ \bibnamefont {Pacold}}, \ and\ \bibinfo {author} {\bibfnamefont {J.~J.}\ \bibnamefont {Rehr}},\ }\href {\doibase 10.1103/PhysRevB.85.115135} {\bibfield  {journal} {\bibinfo  {journal} {Physical Review B}\ }\textbf {\bibinfo {volume} {85}},\ \bibinfo {pages} {115135} (\bibinfo {year} {2012})}\BibitemShut {NoStop}%
\bibitem [{\citenamefont {Klevak}\ \emph {et~al.}(2016)\citenamefont {Klevak}, \citenamefont {Vila}, \citenamefont {Kas}, \citenamefont {Rehr},\ and\ \citenamefont {Seidler}}]{cp_correction}%
  \BibitemOpen
  \bibfield  {author} {\bibinfo {author} {\bibfnamefont {E.}~\bibnamefont {Klevak}}, \bibinfo {author} {\bibfnamefont {F.~D.}\ \bibnamefont {Vila}}, \bibinfo {author} {\bibfnamefont {J.~J.}\ \bibnamefont {Kas}}, \bibinfo {author} {\bibfnamefont {J.~J.}\ \bibnamefont {Rehr}}, \ and\ \bibinfo {author} {\bibfnamefont {G.~T.}\ \bibnamefont {Seidler}},\ }\href {\doibase 10.1103/PhysRevB.94.214201} {\bibfield  {journal} {\bibinfo  {journal} {Physical Review B}\ }\textbf {\bibinfo {volume} {94}},\ \bibinfo {pages} {214201} (\bibinfo {year} {2016})}\BibitemShut {NoStop}%
\bibitem [{\citenamefont {Sternemann}\ \emph {et~al.}(2007)\citenamefont {Sternemann}, \citenamefont {Soininen}, \citenamefont {Sternemann}, \citenamefont {H\"am\"al\"ainen},\ and\ \citenamefont {Tolan}}]{PhysRevB.75.075118}%
  \BibitemOpen
  \bibfield  {author} {\bibinfo {author} {\bibfnamefont {H.}~\bibnamefont {Sternemann}}, \bibinfo {author} {\bibfnamefont {J.~A.}\ \bibnamefont {Soininen}}, \bibinfo {author} {\bibfnamefont {C.}~\bibnamefont {Sternemann}}, \bibinfo {author} {\bibfnamefont {K.}~\bibnamefont {H\"am\"al\"ainen}}, \ and\ \bibinfo {author} {\bibfnamefont {M.}~\bibnamefont {Tolan}},\ }\href {\doibase 10.1103/PhysRevB.75.075118} {\bibfield  {journal} {\bibinfo  {journal} {Physical Review B}\ }\textbf {\bibinfo {volume} {75}},\ \bibinfo {pages} {075118} (\bibinfo {year} {2007})}\BibitemShut {NoStop}%
\bibitem [{\citenamefont {Sternemann}\ \emph {et~al.}(2008)\citenamefont {Sternemann}, \citenamefont {Sternemann}, \citenamefont {Seidler}, \citenamefont {Fister}, \citenamefont {Sakko},\ and\ \citenamefont {Tolan}}]{Sternemann:fh5381}%
  \BibitemOpen
  \bibfield  {author} {\bibinfo {author} {\bibfnamefont {H.}~\bibnamefont {Sternemann}}, \bibinfo {author} {\bibfnamefont {C.}~\bibnamefont {Sternemann}}, \bibinfo {author} {\bibfnamefont {G.~T.}\ \bibnamefont {Seidler}}, \bibinfo {author} {\bibfnamefont {T.~T.}\ \bibnamefont {Fister}}, \bibinfo {author} {\bibfnamefont {A.}~\bibnamefont {Sakko}}, \ and\ \bibinfo {author} {\bibfnamefont {M.}~\bibnamefont {Tolan}},\ }\href {\doibase 10.1107/S0909049508001696} {\bibfield  {journal} {\bibinfo  {journal} {Journal of Synchrotron Radiation}\ }\textbf {\bibinfo {volume} {15}},\ \bibinfo {pages} {162} (\bibinfo {year} {2008})}\BibitemShut {NoStop}%
\bibitem [{\citenamefont {Gatti}\ and\ \citenamefont {Manfredi}(1986)}]{OF_reference}%
  \BibitemOpen
  \bibfield  {author} {\bibinfo {author} {\bibfnamefont {E.}~\bibnamefont {Gatti}}\ and\ \bibinfo {author} {\bibfnamefont {P.~F.}\ \bibnamefont {Manfredi}},\ }\href {\doibase 10.1007/BF02822156} {\bibfield  {journal} {\bibinfo  {journal} {Rivista del Nuovo Cimento}\ }\textbf {\bibinfo {volume} {9N1}},\ \bibinfo {pages} {1} (\bibinfo {year} {1986})}\BibitemShut {NoStop}%
\bibitem [{\citenamefont {Kurinsky}(2018)}]{kurinsky_thesis}%
  \BibitemOpen
  \bibfield  {author} {\bibinfo {author} {\bibfnamefont {N.}~\bibnamefont {Kurinsky}},\ }\href {\doibase https://doi.org/10.2172/1472104} {\emph {\bibinfo {title} {{The low-mass limit: Dark matter detectors with eV-scale energy resolution}}}}\ (\bibinfo  {publisher} {Stanford University},\ \bibinfo {year} {2018})\BibitemShut {NoStop}%
\bibitem [{\citenamefont {Angloher}\ \emph {et~al.}(2023{\natexlab{a}})\citenamefont {Angloher}, \citenamefont {Banik}, \citenamefont {Benato}, \citenamefont {Bento}, \citenamefont {Bertolini}, \citenamefont {Breier}, \citenamefont {Bucci}, \citenamefont {Canonica}, \citenamefont {D'Addabbo}, \citenamefont {Di~Lorenzo} \emph {et~al.}}]{angloher2023latest}%
  \BibitemOpen
  \bibfield  {author} {\bibinfo {author} {\bibfnamefont {G.}~\bibnamefont {Angloher}}, \bibinfo {author} {\bibfnamefont {S.}~\bibnamefont {Banik}}, \bibinfo {author} {\bibfnamefont {G.}~\bibnamefont {Benato}}, \bibinfo {author} {\bibfnamefont {A.}~\bibnamefont {Bento}}, \bibinfo {author} {\bibfnamefont {A.}~\bibnamefont {Bertolini}}, \bibinfo {author} {\bibfnamefont {R.}~\bibnamefont {Breier}}, \bibinfo {author} {\bibfnamefont {C.}~\bibnamefont {Bucci}}, \bibinfo {author} {\bibfnamefont {L.}~\bibnamefont {Canonica}}, \bibinfo {author} {\bibfnamefont {A.}~\bibnamefont {D'Addabbo}}, \bibinfo {author} {\bibfnamefont {S.}~\bibnamefont {Di~Lorenzo}},  \emph {et~al.},\ }\href {\doibase doi = "10.21468/SciPostPhysProc.12.013"} {\bibfield  {journal} {\bibinfo  {journal} {SciPost Physics Proceedings}\ }\textbf {\bibinfo {volume} {12}},\ \bibinfo {pages} {013} (\bibinfo {year} {2023}{\natexlab{a}})}\BibitemShut {NoStop}%
\bibitem [{\citenamefont {Angloher}\ \emph {et~al.}(2023{\natexlab{b}})\citenamefont {Angloher} \emph {et~al.}}]{PhysRevD.107.122003}%
  \BibitemOpen
  \bibfield  {author} {\bibinfo {author} {\bibfnamefont {G.}~\bibnamefont {Angloher}} \emph {et~al.} (\bibinfo {collaboration} {CRESST Collaboration}),\ }\href {\doibase 10.1103/PhysRevD.107.122003} {\bibfield  {journal} {\bibinfo  {journal} {Physical Review D}\ }\textbf {\bibinfo {volume} {107}},\ \bibinfo {pages} {122003} (\bibinfo {year} {2023}{\natexlab{b}})}\BibitemShut {NoStop}%
\bibitem [{\citenamefont {Heikinheimo}\ \emph {et~al.}(2022)\citenamefont {Heikinheimo}, \citenamefont {Sassi}, \citenamefont {Nordlund}, \citenamefont {Tuominen},\ and\ \citenamefont {Mirabolfathi}}]{PhysRevD.106.083009}%
  \BibitemOpen
  \bibfield  {author} {\bibinfo {author} {\bibfnamefont {M.}~\bibnamefont {Heikinheimo}}, \bibinfo {author} {\bibfnamefont {S.}~\bibnamefont {Sassi}}, \bibinfo {author} {\bibfnamefont {K.}~\bibnamefont {Nordlund}}, \bibinfo {author} {\bibfnamefont {K.}~\bibnamefont {Tuominen}}, \ and\ \bibinfo {author} {\bibfnamefont {N.}~\bibnamefont {Mirabolfathi}},\ }\href {\doibase 10.1103/PhysRevD.106.083009} {\bibfield  {journal} {\bibinfo  {journal} {Physical Review D}\ }\textbf {\bibinfo {volume} {106}},\ \bibinfo {pages} {083009} (\bibinfo {year} {2022})}\BibitemShut {NoStop}%
\bibitem [{\citenamefont {Romani}(2024)}]{10.1063/5.0222654}%
  \BibitemOpen
  \bibfield  {author} {\bibinfo {author} {\bibfnamefont {R.~K.}\ \bibnamefont {Romani}},\ }\href {\doibase 10.1063/5.0222654} {\bibfield  {journal} {\bibinfo  {journal} {Journal of Applied Physics}\ }\textbf {\bibinfo {volume} {136}},\ \bibinfo {pages} {124502} (\bibinfo {year} {2024})}\BibitemShut {NoStop}%
\bibitem [{\citenamefont {Anthony-Petersen}\ \emph {et~al.}(2025)\citenamefont {Anthony-Petersen} \emph {et~al.}}]{anthony2024low}%
  \BibitemOpen
  \bibfield  {author} {\bibinfo {author} {\bibfnamefont {R.}~\bibnamefont {Anthony-Petersen}} \emph {et~al.},\ }\href {\doibase 10.1063/5.0247343} {\bibfield  {journal} {\bibinfo  {journal} {Applied Physics Letters}\ }\textbf {\bibinfo {volume} {126}},\ \bibinfo {pages} {102601} (\bibinfo {year} {2025})}\BibitemShut {NoStop}%
\bibitem [{\citenamefont {Alkhatib}\ \emph {et~al.}(2021)\citenamefont {Alkhatib} \emph {et~al.}}]{PhysRevLett.127.061801}%
  \BibitemOpen
  \bibfield  {author} {\bibinfo {author} {\bibfnamefont {I.}~\bibnamefont {Alkhatib}} \emph {et~al.} (\bibinfo {collaboration} {SuperCDMS Collaboration}),\ }\href {\doibase 10.1103/PhysRevLett.127.061801} {\bibfield  {journal} {\bibinfo  {journal} {Physical Review Letter}\ }\textbf {\bibinfo {volume} {127}},\ \bibinfo {pages} {061801} (\bibinfo {year} {2021})}\BibitemShut {NoStop}%
\bibitem [{\citenamefont {Adari}\ \emph {et~al.}(2022)\citenamefont {Adari}, \citenamefont {Aguilar-Arevalo}, \citenamefont {Amidei}, \citenamefont {Angloher}, \citenamefont {Armengaud}, \citenamefont {Augier}, \citenamefont {Balogh}, \citenamefont {Banik}, \citenamefont {Baxter}, \citenamefont {Beaufort} \emph {et~al.}}]{adari2022excess}%
  \BibitemOpen
  \bibfield  {author} {\bibinfo {author} {\bibfnamefont {P.}~\bibnamefont {Adari}}, \bibinfo {author} {\bibfnamefont {A.~A.}\ \bibnamefont {Aguilar-Arevalo}}, \bibinfo {author} {\bibfnamefont {D.}~\bibnamefont {Amidei}}, \bibinfo {author} {\bibfnamefont {G.}~\bibnamefont {Angloher}}, \bibinfo {author} {\bibfnamefont {E.}~\bibnamefont {Armengaud}}, \bibinfo {author} {\bibfnamefont {C.}~\bibnamefont {Augier}}, \bibinfo {author} {\bibfnamefont {L.}~\bibnamefont {Balogh}}, \bibinfo {author} {\bibfnamefont {S.}~\bibnamefont {Banik}}, \bibinfo {author} {\bibfnamefont {D.}~\bibnamefont {Baxter}}, \bibinfo {author} {\bibfnamefont {C.}~\bibnamefont {Beaufort}},  \emph {et~al.},\ }\href {\doibase 10.21468/SciPostPhysProc.9.001} {\bibfield  {journal} {\bibinfo  {journal} {SciPost Physics Proceedings}\ }\textbf {\bibinfo {volume} {9}},\ \bibinfo {pages} {001} (\bibinfo {year} {2022})}\BibitemShut {NoStop}%
\bibitem [{\citenamefont {Cowan}(1998)}]{Cowan:1998ji}%
  \BibitemOpen
  \bibfield  {author} {\bibinfo {author} {\bibfnamefont {G.}~\bibnamefont {Cowan}},\ }\href {https://global.oup.com/academic/product/statistical-data-analysis-9780198501565} {\emph {\bibinfo {title} {Statistical Data Analysis}}}\ (\bibinfo  {publisher} {Oxford University Press, USA},\ \bibinfo {year} {1998})\ \bibinfo {note} {\href{https://global.oup.com/academic/product/statistical-data-analysis-9780198501565}{Oxford University Press}}\BibitemShut {NoStop}%
\bibitem [{\citenamefont {Ramanathan}\ and\ \citenamefont {Kurinsky}(2020)}]{eh_pair_creation_prob}%
  \BibitemOpen
  \bibfield  {author} {\bibinfo {author} {\bibfnamefont {K.}~\bibnamefont {Ramanathan}}\ and\ \bibinfo {author} {\bibfnamefont {N.}~\bibnamefont {Kurinsky}},\ }\href {\doibase 10.1103/PhysRevD.102.063026} {\bibfield  {journal} {\bibinfo  {journal} {Physical Review D}\ }\textbf {\bibinfo {volume} {102}},\ \bibinfo {pages} {063026} (\bibinfo {year} {2020})}\BibitemShut {NoStop}%
\bibitem [{\citenamefont {Neganov}\ and\ \citenamefont {Trofimov}(1978)}]{neganov1978possibility}%
  \BibitemOpen
  \bibfield  {author} {\bibinfo {author} {\bibfnamefont {B.~S.}\ \bibnamefont {Neganov}}\ and\ \bibinfo {author} {\bibfnamefont {V.~N.}\ \bibnamefont {Trofimov}},\ }\href {https://www.osti.gov/biblio/6230258} {\bibfield  {journal} {\bibinfo  {journal} {Journal of Experimental and Theoretical Physics Letters}\ }\textbf {\bibinfo {volume} {28}},\ \bibinfo {pages} {328} (\bibinfo {year} {1978})},\ \bibinfo {note} {\href{https://www.osti.gov/biblio/6230258}{OSTI Record}}\BibitemShut {NoStop}%
\bibitem [{\citenamefont {Luke}(1988)}]{10.1063/1.341976}%
  \BibitemOpen
  \bibfield  {author} {\bibinfo {author} {\bibfnamefont {P.~N.}\ \bibnamefont {Luke}},\ }\href {\doibase 10.1063/1.341976} {\bibfield  {journal} {\bibinfo  {journal} {Journal of Applied Physics}\ }\textbf {\bibinfo {volume} {64}},\ \bibinfo {pages} {6858} (\bibinfo {year} {1988})}\BibitemShut {NoStop}%
\bibitem [{\citenamefont {Kelsey}\ \emph {et~al.}(2023)\citenamefont {Kelsey} \emph {et~al.}}]{geant4cmp}%
  \BibitemOpen
  \bibfield  {author} {\bibinfo {author} {\bibfnamefont {M.}~\bibnamefont {Kelsey}} \emph {et~al.},\ }\href {\doibase 10.1016/j.nima.2023.168473} {\bibfield  {journal} {\bibinfo  {journal} {Nuclear Instruments and Methods in Physics Research Section A: Accelerators, Spectrometers, Detectors and Associated Equipment}\ }\textbf {\bibinfo {volume} {1055}},\ \bibinfo {pages} {168473} (\bibinfo {year} {2023})}\BibitemShut {NoStop}%
\bibitem [{\citenamefont {Kelsey}\ \emph {et~al.}(2022)\citenamefont {Kelsey} \emph {et~al.}}]{geant4cmp_code}%
  \BibitemOpen
  \bibfield  {author} {\bibinfo {author} {\bibfnamefont {M.}~\bibnamefont {Kelsey}} \emph {et~al.},\ }\href@noop {} {\enquote {\bibinfo {title} {{G4CMP} code repository},}\ }\bibinfo {howpublished} {\url{https://github.com/kelseymh/G4CMP}} (\bibinfo {year} {2022})\BibitemShut {NoStop}%
\end{thebibliography}%

\end{document}